\documentclass[journal,12pt,onecolumn,draftclsnofoot,]{IEEEtran}
\usepackage[utf8]{inputenc}
\usepackage[T1]{fontenc}
\usepackage{amsmath,amsfonts}
\usepackage{mathtools}
\usepackage{algorithm}
\usepackage{algorithmic}
\usepackage{cite}
\usepackage{graphicx}
\usepackage[caption=false, font=footnotesize]{subfig}
\usepackage{tikz}
\usetikzlibrary{arrows,shapes,positioning,patterns,decorations.pathreplacing}

\newtheorem{theorem}{Theorem}
\newtheorem{definition}{Definition}
\newtheorem{lemma}{Lemma}
\newtheorem{proposition}{Proposition}

\newcommand{\Prob}{\mathsf{Pr}}
\newcommand{\E}{\operatorname{E}}

\title{Structural Complexity of One-Dimensional Random Geometric Graphs} 
\author{Mihai-Alin~Badiu~\IEEEmembership{Member,~IEEE} and Justin~P.~Coon,~\IEEEmembership{Senior Member,~IEEE}
\thanks{The authors are with the Department of Engineering Science, University of Oxford, Parks Road, Oxford, OX1 3PJ, UK, (email: mihai.badiu@eng.ox.ac.uk; justin.coon@eng.ox.ac.uk).}
\thanks{This material is based upon work supported by, or in part by, the U. S. Army Research Laboratory and the U. S. Army Research Office under contract/grant number W911NF-19-1-0048, and the EPSRC grant number EP/T02612X/1.}
\thanks{For the purpose of open access, the author(s) has applied a Creative Commons Attribution (CC BY) license to any Accepted Manuscript version arising.}
}

\begin{document}

\maketitle

\begin{abstract}
We study the richness of the ensemble of graphical structures (i.e., unlabeled graphs) of the one-dimensional random geometric graph model defined by $n$ nodes randomly scattered in $[0,1]$ that connect if they are within the connection range $r\in[0,1]$.
We provide bounds on the number of possible structures which give universal upper bounds on the structural entropy that hold for any $n$, $r$ and distribution of the node locations. For fixed $r$, the number of structures is $\Theta(a^{2n})$ with $a=a(r)=2 \cos{\left(\frac{\pi}{\lceil 1/r \rceil+2}\right)}$, and therefore the structural entropy is upper bounded by $2n\log_2 a(r) + O(1)$. For large $n$, we derive bounds on the structural entropy normalized by $n$, and evaluate them for independent and uniformly distributed node locations. When the connection range $r_n$ is $O(1/n)$, the obtained upper bound is given in terms of a function that increases with $n r_n$  and asymptotically attains $2$ bits per node. If the connection range is bounded away from zero and one, the upper and lower bounds decrease linearly with $r$, as $2(1-r)$ and $(1-r)\log_2 e$, respectively. When $r_n$ is vanishing but dominates $1/n$ (e.g., $r_n \propto \ln n / n$), the normalized entropy is between $\log_2 e \approx 1.44$ and $2$ bits per node. We also give a simple encoding scheme for random structures that requires $2$ bits per node. The upper bounds in this paper easily extend to the entropy of the \emph{labeled} random graph model, since this is given by the structural entropy plus a term that accounts for all the permutations of node labels that are possible for a given structure, which is no larger than $\log_2(n!) = n \log_2 n - n + O(\log_2 n)$.
\end{abstract}


\section{Introduction}
Graphs are important mathematical structures used in network science for representing and analysing systems composed of multiple entities (represented by the nodes of the graph) that are related or interact in some sense (indicated through edges that connect some pairs of nodes). There are a multitude of applications in various disciplines, including biological systems (e.g., protein-protein interactions, neuronal networks), information networks (computer networks, wireless sensor networks), social networks, economic networks, etc. 

Several random graph models whose typical instances are representative of real-world networks exist in the literature~\cite{Bollobas2001,Newman2010}. The Erd\H{o}s-R\'{e}nyi model where edges between $n$ nodes are drawn independently with probability $p$ is an emblematic basic model that is convenient for theoretical analysis, although it does not exhibit important properties of real networks, such as transitivity, clustering, or inhomogeneity. 

As many real networks are increasingly large, the storage or transmission of graphs and data on graphs becomes problematic. It thus becomes important to quantify the amount of information embodied in networks and develop efficient methods of representing graphs and data associated with them compactly by exploiting any potential statistical redundancy. From an information theoretical perspective, the limit to compression of random graphs is the entropy of the graph distribution, according to Shannon's source coding theorem. The case of the Erd\H{o}s-R\'{e}nyi model is simple, as the random graph can be treated as a collection of ${n \choose 2}$ independent binary sources, such that the entropy is ${n \choose 2} h(p)$ with $h(p) = -p\log_2 p - (1-p)\log_2(1-p)$. More realistic random graphs, however, exhibit intricate dependencies which makes exact computation complicated.

Graphical data are structurally different from the conventional linearly ordered arrays (time series or higher-dimensional data~\cite{Lempel1986}), in that they have an intrinsic graph structure. A relevant question is how much of the complexity of a graph is structural, i.e., pertaining to the unlabeled graph, and how much of it is data complexity related to the labels carried by the nodes for given structure (equivalently, how informative the random structure is about the labels of the graph). Furthermore, there are cases in which it is not necessary to distinguish between the nodes, i.e., their identities are irrelevant. For example, many of the graph properties (e.g., node degrees, connectedness, $k$-connectivity, etc.) depend on the structures in the ensemble, and not on the labels of the nodes. In such cases, one may be interested in storing or transmitting only the structural information relevant for reconstructing the unlabelled graph (i.e., up to graph isomorphisms). That is, the focus is on the graph structures and the problem is of compressing graphs with node labels removed. 

Compression of unlabeled graphs was addressed by Tur\'{a}n~\cite{Turan1984} more than three decades ago, who showed that unlabeled planar graphs can be represented (optimally within a constant factor) using $12n$ bits and gave an asymptotically optimal encoding of labeled planar graphs that requires $n\left \lceil{\log_2 n}\right \rceil + 12n$ bits. Tur\'{a}n also suggested a lower bound on the number of bits in the representation of general unlabeled graphs of ${n \choose 2} - n\log_2 n + O(n)$ and raised the problem of finding an efficient (i.e., computable in polynomial time) coding method. This problem was solved a few years later by Naor~\cite{Naor1990}, who proposed a linear-time encoding method that is optimal up to the $O(n)$ term. More recently, Choi \& Szpankowski~\cite{Choi2009,Choi2012} addressed compression of unlabeled graphs (which they name \emph{graphical structures}) for the Erd\H{o}s-R\'{e}nyi model and showed that the corresponding Shannon entropy (coined \emph{structural entropy}) is ${n \choose 2}h(p) - n\log_2 n + O(n)$. Note the $n\log_2 n$ reduction of the entropy compared to the entropy of labeled Erd\H{o}s-R\'{e}nyi graphs. A similar study of entropy as a measure of structural complexity is performed in~\cite{Kieffer2009} for random binary trees. Motivated by the fact that real networks are inhomogeneous and exhibit cluster structures, properties that are not captured by the Erd\H{o}s-R\'{e}nyi model,~\cite{Abbe2016} studies fundamental limits on the compression of the stochastic block model (SBM) with labeled vertices,~\cite{Bhatt2021} proposes a universal graph compressor for SBM, whereas~\cite{Asadi17} investigates limits for compressing data on SBM graphs.
The structural entropy and the compression of preferential attachment graphs are addressed in~\cite{Luczak2019asymmetry,Luczak2019}, while~\cite{Delgosha2021} proposes a compression method for sparse graphs and heavy-tailed sparse graphs.

Since many real-world networks are spatial~\cite{Barthelemy2011}, the random geometric graph (RGG, where pairs of nodes are connected based on the distance between them in some latent space~\cite{Penrose2003}) is often used as a model of practical applications as it captures well the characteristics of information networks, biological systems, social networks, and economic networks. The entropy of spatial graph models was studied in~\cite{Coon2016,Coon2018,Badiu2018}, where simple bounds on the entropy of labelled graphs were obtained based on the subadditivity property of joint entropy; the bounds scale as $n^2$, as in the case of the Erd\H{o}s-R\'{e}nyi graph model. However, RGGs have salient features (e.g., strong transitivity) induced by the spatial embedding of the nodes and distance-based connectivity that have not been fully exploited in the aforementioned bounds. This motivates us to investigate the structural complexity of RGGs more closely.

In this paper, we study how rich the ensemble of structures (i.e., unlabeled graphs) of the 1D RGG model is, how its complexity (quantified by the structural entropy) scales with the number of nodes and how the  connection range impacts the structural entropy. In Sec.~\ref{sec:prelims}, we define the considered 1D RGG model, which has $n$ nodes randomly located on $[0,1]$ and connection range $r$. We also discuss its connectivity and introduce the notions of graphical structures and structural entropy. 

In Sec.~\ref{sec:struct}, we make a connection between the structure of the 1D RGG and a particular graph, which we call the \emph{ordered graph}, whose nodes are indexed according to the order of their underlying locations. We show that, up to left-right reversal, the ordered graphs identify the connected graphical structures produced by the 1D RGG model, and the total number of possible structures in the model is upper bounded by the number of ordered graphs. This enables us to characterize in Sec.~\ref{sec:count} the structural complexity of the 1D RGG model by estimating the number of possible ordered graphs. Based on their representation in terms of maximal cliques, we find that the number of ordered graphs and consequently the number of structures with $n$ nodes is upper limited by the $n$th Catalan number, whereas the number of connected structures is bounded by the $(n-1)$th Catalan number. We also identify an interesting link between ordered graphs and existing combinatorial objects known as Dyck paths. However, the obtained upper bound does not depend on the connection range $r$, whereas the value of $r$ restricts the number of possible ordered graphs and structures. We then improve upon this result by finding a way to express the exact number of ordered graphs and determining its generating function. Interestingly, the obtained generating function corresponds to a known sequence that gives the number of certain combinatorial objects, including height-restricted Dyck paths and ordered rooted trees. In a similar manner, we additionally find a lower bound on the number of connected ordered graphs, and then obtain that for fixed $r$ the number of structures is $\Theta(a^{2n})$ with $a=a(r)=2 \cos{\left(\frac{\pi}{\lceil 1/r \rceil+2}\right)}$. Based on this count, in Sec.~\ref{sec:struct_ent} we immediately obtain an upper bound on the structural entropy (Theorem~\ref{th:UBcount}), which is asymptotically attained for fixed $r$ and uniform distribution of the structures.

We subsequently establish bounds on the structural entropy in terms of the entropy of the ordered graph. A particularly relevant result is that for connected graphs, the entropy of the ordered graph determines the structural entropy within one bit; this is important, because for independent and uniformly distributed (i.u.d.) points the RGG is connected with probability one, as $n\to\infty$, when the connection range $r_n$ is above the critical threshold $\ln(n)/n$. Thus, the entropy of the ordered graph is an asymptotically precise representation of the structural entropy in that regime. We evaluate the bounds for i.u.d. point locations (which is typically assumed for the 1D RGG), for different scaling regimes of the connection range $r$, as $n$ grows large (Theorem~\ref{th:iud}). In particular, in the asymptotically connected regime, which is of practical interest and relevance, we find that the normalized structural entropy is somewhere between $\log_2 e \approx 1.44$ and $2$ bits per node, when $r_n$ is vanishing, and it is between $\log_2 e (1-r)$ and $2(1-r)$ for fixed $r$. We also give a simple encoding scheme for 1D random geometric graphical structures that requires $2n$ bits, i.e., it achieves the obtained upper limit in the asymptotically connected regime. 

The upper bounds in this paper easily extend to the entropy of the \emph{labeled} 1D RGG, since this is no larger than the structural entropy plus a term that accounts for all the permutations of node labels that are possible for a given structure, which is no larger than $\log_2(n!) = n \log_2 n - n + O(\log_2 n)$.

\section{Preliminaries}\label{sec:prelims}
\subsection{Graph Model}\label{sec:model}
The one-dimensional random geometric graph $G_{n,r}$ is constructed by placing $n$ points randomly on a line segment and connecting by an edge every pair of nodes that are separated by a distance smaller than a threshold $r$. More formally, assume the $n$ points have the random locations $X_1,\ldots,X_n$ in the interval $[0,1]$. For now, we refrain from defining the (joint) distribution of $X_1,\ldots,X_n$, since some results shown below are universal. However, we consider only distributions such that each point can belong to any subinterval of $[0,1]$ with nonzero probability. We will consider the specific case where $X_1,\ldots,X_n$ are i.u.d.\@ in Theorem~\ref{th:iud}.  The corresponding graph $G_{n,r} = (V,E)$ has the set of vertices  $V=\{1,\ldots,n\}$ and the set of undirected edges $E=\{(i,j)\in V^2 \mid |X_i-X_j|\leq r,\, i < j \}$, for a fixed connection range $r>0$. 

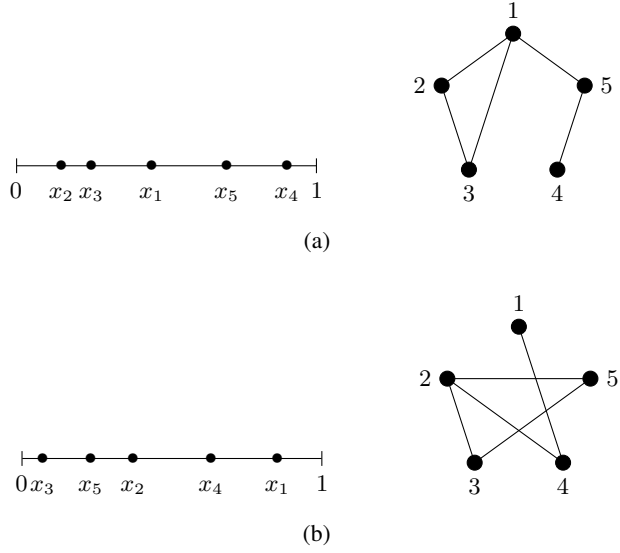
\begin{figure}
    \centering
    \subfloat[\label{fig:example_RGG_a}]{%
        \begin{tikzpicture}
            \footnotesize
            \draw[|-|] (0,0) -- (4,0)
            node[pos=0,label=below:{$0$}]{}
            node[pos=1,label=below:{$1$}]{}
            node[pos=0.15,label=below:{$x_2$}]{$\bullet$}
            node[pos=0.25,label=below:{$x_3$}]{$\bullet$}
            node[pos=0.45,label=below:{$x_1$}]{$\bullet$}
            node[pos=0.7,label=below:{$x_5$}]{$\bullet$}
            node[pos=0.9,label=below:{$x_4$}]{$\bullet$};
        \end{tikzpicture}
        \qquad
        \begin{tikzpicture}
            \footnotesize
            \node[fill,circle,inner sep=2pt,outer sep=0pt,draw,label=above:$1$](N1) at (1*360/5+18:1) {};
            \node[fill,circle,inner sep=2pt,outer sep=0pt,draw,label=left:$2$](N2) at (2*360/5+18:1) {};
            \node[fill,circle,inner sep=2pt,outer sep=0pt,draw,label=below:$3$](N3) at (3*360/5+18:1) {};
            \node[fill,circle,inner sep=2pt,outer sep=0pt,draw,label=below:$4$](N4) at (4*360/5+18:1) {};
            \node[fill,circle,inner sep=2pt,outer sep=0pt,draw,label=right:$5$](N5) at (5*360/5+18:1) {};
            \draw (N1) -- (N2);
            \draw (N1) -- (N3);
            \draw (N1) -- (N5);
            \draw (N2) -- (N3);
            \draw (N4) -- (N5);
        \end{tikzpicture} }
    \\
    \subfloat[\label{fig:example_RGG_b}]{%
        \begin{tikzpicture}
            \footnotesize
            \draw[|-|] (0,0) -- (4,0)
            node[pos=0,label=below:{$0$}]{}
            node[pos=1,label=below:{$1$}]{}
            node[pos=0.07,label=below:{$x_3$}]{$\bullet$}
            node[pos=0.23,label=below:{$x_5$}]{$\bullet$}
            node[pos=0.37,label=below:{$x_2$}]{$\bullet$}
            node[pos=0.63,label=below:{$x_4$}]{$\bullet$}
            node[pos=0.85,label=below:{$x_1$}]{$\bullet$};
        \end{tikzpicture}
        \qquad
        \begin{tikzpicture}
            \footnotesize
            \node[fill,circle,inner sep=2pt,outer sep=0pt,draw,label=above:$1$](N1) at (1*360/5+18:1) {};
            \node[fill,circle,inner sep=2pt,outer sep=0pt,draw,label=left:$2$](N2) at (2*360/5+18:1) {};
            \node[fill,circle,inner sep=2pt,outer sep=0pt,draw,label=below:$3$](N3) at (3*360/5+18:1) {};
            \node[fill,circle,inner sep=2pt,outer sep=0pt,draw,label=below:$4$](N4) at (4*360/5+18:1) {};
            \node[fill,circle,inner sep=2pt,outer sep=0pt,draw,label=right:$5$](N5) at (5*360/5+18:1) {};
            \draw (N1) -- (N4);
            \draw (N2) -- (N3);
            \draw (N2) -- (N4);
            \draw (N2) -- (N5);
            \draw (N3) -- (N5);
        \end{tikzpicture}}
  \caption{Two outcomes of $G_{n,r}$ for $n=5$ and $r=1/3$; the  points are i.u.d. on $[0,1]$ (left); the resulting graph (right).}
  \label{fig:example_RGG} 
\end{figure}

The graph $G_{n,r}$ depends deterministically on the node locations. However, it does not retain the locations but only the connectivity among the nodes. Also, an infinite number of configurations of locations may be mapped to the same graph. We denote by $\mathcal{G}_{n,r}$ the set of all possible graphs that are produced by this model, i.e., the support of $G_{n,r}$. Fig.~\ref{fig:example_RGG} illustrates two such graphs. The probability distribution of $(X_1,\ldots,X_n)$ induces a distribution of graphs in $\mathcal{G}_{n,r}$, such that a graph $G\in\mathcal{G}_{n,r}$ occurs with probability $P(G)$.

\subsection{Connectivity}\label{sec:conn}
A graph is said to be connected if there exists at least one path between every pair of distinct vertices. The connectivity of one-dimensional random geometric graphs has been studied extensively, e.g., see \cite{Han2007} and the references therein. Trivially, $G_{n,r}$ is connected for $r\geq 1$, as $G_{n,r}$ is deterministically a complete graph in that case. For $r\in(0,1)$ and finite $n$, the graph is connected with a probability smaller than one in general (e.g., when the random locations have a non-vanishing density on $[0,1]$); for $X_1,\ldots,X_n$ independent and uniformly distributed, the probability is available in a closed form expression~\cite{Desai2002,Ghasemi2006}. 

Most of the works consider the connectivity of $G_{n,r}$ for large $n$ and study how the connection range $r$ should scale with $n$ to achieve connectivity, i.e., what scaling functions $r:\mathbb{N}\to \mathbb{R}_+$, $n\to r_n$, are appropriate. For i.u.d. point locations, the connectivity exhibits a typical behaviour in the sense that there exists a critical range function $r^\ast_n = \frac{\ln n}{n}$ such that, as $n$ becomes large, the graph is connected (or disconnected) with high probability depending on how the scaling $r$ that is being used deviates from the critical scaling $r^\ast$. Specifically, for a connection range function in the form $r_n = r_n^\ast + \frac{\alpha_n}{n}$ for some $\alpha:\mathbb{N}\to \mathbb{R}$, it holds that\cite{Han2007}
\begin{equation}\label{eq:Pcon}
     \lim_{n\to\infty} \Prob(G_{n,r}\text{ is connected}) =
    \begin{cases}
    0, &\text{if } \lim_{n\to\infty} \alpha_n = -\infty,\\
    1, &\text{if } \lim_{n\to\infty} \alpha_n = \infty.
    \end{cases}
\end{equation}
For example, a deviation function $\alpha_n = \pm \ln\ln n$ determines $G_{n,r}$ to be connected or disconnected with probability one depending on the sign, in the limit of large $n$. More generally, it is shown in~\cite{Han2009} that graph connectivity also obeys a strong zero-one law when the distribution of the point locations has a non-vanishing density. 

\subsection{Graphical Structures}
The graphical structures produced by the model of Sec.~\ref{sec:model} can be defined formally based on the notion of graph isomorphism. 
\begin{definition}[Structure]
Two graphs $G_1 = (V,E_1)\in\mathcal{G}_{n,r}$ and $G_2 = (V,E_2)\in\mathcal{G}_{n,r}$ have the same structure, which is denoted as $G_1\simeq G_2$, if and only if there exists a graph isomorphism between them, i.e., if and only if there exists an adjacency-preserving permutation of the vertices $\pi:V\to V$ such that $(i,j)\in E_1 \iff (\pi(i),\pi(j))\in E_2$, for all $i,j\in V$.
\end{definition}

For example, one can verify that the two graphs in Fig.~\ref{fig:example_RGG} have the same structure, because the permutation $\pi(1)=2$, $\pi(2)=3$, $\pi(3)=5$, $\pi(4)=1$ and $\pi(5)=4$ (applied to the vertices in Fig.~\ref{fig:example_RGG_a}) is edge preserving. 

The relation `$\simeq$' is an equivalence relation, which partitions the set $\mathcal{G}_{n,r}$ into disjoint equivalence classes. All graphs in an equivalence class have the same structure, while graphs in different classes are structurally different. Thus, one can identify a structure by its defining equivalence class or by a representative member of that class. In the following, we denote by $\mathcal{S}_{n,r}$ the set of all structures of $\mathcal{G}_{n,r}$.

\subsection{Structural Entropy}
The probability distribution of $G_{n,r}$ with pmf $P(G)$, $G\in\mathcal{G}_{n,r}$, induces a probability distribution over the set of structures $\mathcal{S}_{n,r}$. Let $f:\mathcal{G}_{n,r} \to \mathcal{S}_{n,r}$ be a deterministic surjection that maps each graph in $\mathcal{G}_{n,r}$ to its corresponding structure in $\mathcal{S}_{n,r}$. Accordingly, the preimage $f^{-1}(S)$ of a structure $S\in\mathcal{S}_{n,r}$ is the set of all graphs in $\mathcal{G}_{n,r}$ that belong to the equivalent class represented by $S$.  Based on the distribution of graphs, the probability of a structure $S\in\mathcal{S}_{n,r}$ can thus be expressed as
\begin{equation} \label{eq:Ps}
    P(S) = \sum_{G \in f^{-1}(S)} P(G).
\end{equation}
The entropy of $G_{n,r}$ is
\begin{equation}
    H(G_{n,r}) = - \sum_{G\in\mathcal{G}_{n,r}} P(G) \log_2 P(G)
\end{equation}
whereas the structural entropy is defined by~\cite{Choi2012}
\begin{equation}\label{eq:struct_entropy_def}
    H(S_{n,r}) = - \sum_{S\in\mathcal{S}_{n,r}} P(S) \log_2 P(S)
\end{equation}

\section{Structures of $\mathcal{G}_{n,r}$}\label{sec:struct}
\subsection{Graph Ordering}
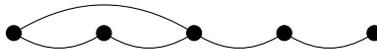
\begin{figure}
    \centering
    \begin{tikzpicture}
        \node[fill,circle,inner sep=2pt,outer sep=0pt,draw](A){};
        \node[fill,circle,inner sep=2pt,outer sep=0pt,draw,right=of A](B){};
        \node[fill,circle,inner sep=2pt,outer sep=0pt,draw,right=of B](C){};
        \node[fill,circle,inner sep=2pt,outer sep=0pt,draw,right=of C](D){};
        \node[fill,circle,inner sep=2pt,outer sep=0pt,draw,right=of D](E){};
        \draw[bend right](A) to node[]{} (B);
        \draw[bend left](A) to node[]{} (C);
        \draw[bend right](B) to node[]{} (C);
        \draw[bend right](C) to node[]{} (D);
        \draw[bend right](D) to node[]{} (E);
    \end{tikzpicture}
\caption{The common structure of the graphs depicted in Fig.~\ref{fig:example_RGG}. The nodes are displayed in the order of the point locations and their labels are discarded.}
\label{fig:example_struct}
\end{figure}

A one-dimensional random geometric graph is naturally visualised by depicting its nodes in the order of their locations. For example, by doing so, we realize that the graphs in Fig.~\ref{fig:example_RGG_a} and~\ref{fig:example_RGG_b} have the same structure, which is illustrated in Fig.~\ref{fig:example_struct}. In the following such a representation of structure is formally justified. For connected structures, we show that each equivalence class can be represented by a particular graph whose nodes are indexed according to the order of the underlying point locations. This special graph, which we call `ordered graph', is then designated as the structure representative. Disconnected structures can be treated as a collection of smaller, connected structures.

\begin{definition}[Ordered graph]
Let $X_1,\ldots,X_n \in [0,1]$ be the random locations of $n$ points and let $X_{(1)},\ldots,X_{(n)}$ be the order statistics of the $n$ locations, i.e., $X_{(k)}$ is the $k$th smallest of the $X_1,\ldots,X_n$. The random graph $G_{n,r}$ is constructed as in Sec.~\ref{sec:model}. We define the ordered graph $\hat G_{n,r}$ as the graph with nodes $ V=\{1,\ldots,n\}$ and edges $\hat E=\{(i,j)\in V\times V\mid |X_{(i)}-X_{(j)}|\leq r, i < j\}$. 
\end{definition}

In the following, $\hat{\mathcal{G}}_{n,r}$ denotes the support of the random ordered graph $\hat{G}_{n,r}$.
\begin{lemma}\label{lemma:OG_G_iso}
The graphs $G_{n,r}$ and $\hat G_{n,r}$ are isomorphic.
\end{lemma}
\begin{IEEEproof}
Let $\sigma:\{1,\ldots,n\} \to \{1,\ldots,n\}$ be the bijection that reindexes the points in ascending order of their location, such that $X_{\sigma^{-1}(1)} \leq X_{\sigma^{-1}(2)} \leq \ldots \leq X_{\sigma^{-1}(n)}$, i.e., $X_{\sigma^{-1}(i)} \equiv X_{(i)}$ is the $i$th order statistic of the random locations. From the definitions of $G_{n,r}$ and $\hat G_{n,r}$, it can be verified that $\sigma$ is an edge-preserving permutation and therefore an isomorphism.
\end{IEEEproof}


\subsection{Maximal Cliques} \label{sec:max_cliques}
In the following, an ordered geometric graph is represented in terms of its constituent maximal cliques. A clique of a graph is a subset of the vertices of the graph, such that any two vertices are adjacent, i.e., the subgraph induced by the clique is complete. A clique is maximal if it is not a subset of a larger clique, i.e., no adjacent vertex can be included to extend the clique. The number of maximal cliques of a graph with $n$ vertices can take any value between one and $n$, the extremes corresponding to the complete graph and respectively the empty graph (in which case each maximal clique consists of one isolated node).

Ordered graphs have the following basic property. 
\begin{lemma}\label{lemma:OG_clique}
If nodes $i$ and $j$, $i<j$, of $\hat G_{n,r}$ are connected by an edge, then the set of vertices $\{i,i+1,\ldots,j\}$ is a clique. 
\end{lemma} 
\begin{IEEEproof}
Given that $(i,j)$ is an edge, $|X_{(j)}-X_{(i)}|\leq r$. Furthermore, for the order statistic $(X_{(1)},\ldots,X_{(n)})$, it holds that $|X_{(l)}-X_{(k)}|\leq |X_{(j)}-X_{(i)}|\leq r$, for all $k,l\in \{i,i+1,\ldots,j\}$. It follows that $(k,l)$ is an edge, for every $k,l\in \{i,i+1,\ldots,j\}$, and therefore $\{i,i+1,\ldots,j\}$ is a clique.
\end{IEEEproof}

\begin{figure}
    \centering
    \subfloat[\label{fig:locations}]{%
        \begin{tikzpicture}
            \footnotesize
            \draw[|-|] (0,0) -- (6,0)
            node[pos=0,label=below:{$0$}]{}
            node[pos=1,label=below:{$1$}]{}
            node[pos=0.11]{$\bullet$}
            node[pos=0.19]{$\bullet$}
            node[pos=0.24]{$\bullet$}
            node[pos=0.29]{$\bullet$}
            node[pos=0.32]{$\bullet$}
            node[pos=0.46]{$\bullet$}
            node[pos=0.51]{$\bullet$}
            node[pos=0.73]{$\bullet$}
            node[pos=0.76]{$\bullet$}
            node[pos=0.83]{$\bullet$};
        \end{tikzpicture}}
    \\
    \subfloat[\label{fig:ordGraph}]{%
        \def\d{0.8}
        \begin{tikzpicture}
            \footnotesize
            \node[fill,circle,inner sep=2pt,outer sep=0pt,draw,label=below:$1$] (N1)   at (0*\d,0) {};
            \node[fill,circle,inner sep=2pt,outer sep=0pt,draw,label=below:$2$] (N2)   at (1*\d,0) {};
            \node[fill,circle,inner sep=2pt,outer sep=0pt,draw,label=below:$3$] (N3)   at (2*\d,0) {};
            \node[fill,circle,inner sep=2pt,outer sep=0pt,draw,label=below:$4$] (N4)   at (3*\d,0) {};
            \node[fill,circle,inner sep=2pt,outer sep=0pt,draw,label=below:$5$] (N5)   at (4*\d,0) {};
            \node[fill,circle,inner sep=2pt,outer sep=0pt,draw,label=below:$6$] (N6)   at (5*\d,0) {};
            \node[fill,circle,inner sep=2pt,outer sep=0pt,draw,label=below:$7$] (N7)   at (6*\d,0) {};
            \node[fill,circle,inner sep=2pt,outer sep=0pt,draw,label=below:$8$] (N8)   at (7*\d,0) {};
            \node[fill,circle,inner sep=2pt,outer sep=0pt,draw,label=below:$9$] (N9)   at (8*\d,0) {};
            \node[fill,circle,inner sep=2pt,outer sep=0pt,draw,label=below:$10$](N10)  at (9*\d,0) {};
            \draw[bend right](N1) to node[]{} (N2);
            \draw[bend left] (N1) to node[]{} (N3);
            \draw[bend left] (N1) to node[]{} (N4);
            \draw[bend right](N2) to node[]{} (N3);
            \draw[bend left] (N2) to node[]{} (N4);
            \draw[bend left] (N2) to node[]{} (N5);
            \draw[bend right](N3) to node[]{} (N4);
            \draw[bend left] (N3) to node[]{} (N5);
            \draw[bend right](N4) to node[]{} (N5);
            \draw[bend left] (N4) to node[]{} (N6);
            \draw[bend right](N5) to node[]{} (N6);
            \draw[bend left] (N5) to node[]{} (N7);
            \draw[bend right](N6) to node[]{} (N7);
            \draw[bend right](N8) to node[]{} (N9);
            \draw[bend left](N8) to node[]{} (N10);
            \draw[bend right](N9) to node[]{} (N10);
    \end{tikzpicture}}
    \\
    \subfloat[\label{fig:decomp}]{%
        \def\d{0.8}
        \begin{tikzpicture}
            \footnotesize
            \node[fill,circle,inner sep=2pt,outer sep=0pt,draw,label=below:$1$] (N11)   at (0*\d,-0*\d) {};
            \node[fill,circle,inner sep=2pt,outer sep=0pt,draw,label=below:$2$] (N21)   at (1*\d,-0*\d) {};
            \node[fill,circle,inner sep=2pt,outer sep=0pt,draw,label=below:$3$] (N31)   at (2*\d,-0*\d) {};
            \node[fill,circle,inner sep=2pt,outer sep=0pt,draw,label=below:$4$] (N41)   at (3*\d,-0*\d) {};
            \node[fill,circle,inner sep=2pt,outer sep=0pt,draw,label=below:$2$] (N22)   at (1*\d,-1*\d) {};
            \node[fill,circle,inner sep=2pt,outer sep=0pt,draw,label=below:$3$] (N32)   at (2*\d,-1*\d) {};
            \node[fill,circle,inner sep=2pt,outer sep=0pt,draw,label=below:$4$] (N42)   at (3*\d,-1*\d) {};
            \node[fill,circle,inner sep=2pt,outer sep=0pt,draw,label=below:$5$] (N52)   at (4*\d,-1*\d) {};
            \node[fill,circle,inner sep=2pt,outer sep=0pt,draw,label=below:$4$] (N43)   at (3*\d,-2*\d) {};
            \node[fill,circle,inner sep=2pt,outer sep=0pt,draw,label=below:$5$] (N53)   at (4*\d,-2*\d) {};
            \node[fill,circle,inner sep=2pt,outer sep=0pt,draw,label=below:$6$] (N63)   at (5*\d,-2*\d) {};
            \node[fill,circle,inner sep=2pt,outer sep=0pt,draw,label=below:$5$] (N54)   at (4*\d,-3*\d) {};
            \node[fill,circle,inner sep=2pt,outer sep=0pt,draw,label=below:$6$] (N64)   at (5*\d,-3*\d) {};
            \node[fill,circle,inner sep=2pt,outer sep=0pt,draw,label=below:$7$] (N74)   at (6*\d,-3*\d) {};
            \node[fill,circle,inner sep=2pt,outer sep=0pt,draw,label=below:$8$] (N85)   at (7*\d,-4*\d) {};
            \node[fill,circle,inner sep=2pt,outer sep=0pt,draw,label=below:$9$] (N95)   at (8*\d,-4*\d) {};
            \node[fill,circle,inner sep=2pt,outer sep=0pt,draw,label=below:$10$](N105)  at (9*\d,-4*\d) {};
        \end{tikzpicture}}
    \caption{Illustration of the decomposition of an ordered graph into maximal cliques: (a) the locations of $n=10$ points (b) the ordered graph determined by the set of locations and connection range $r=0.2$; (c) the ordered graph consists of five maximal cliques: $[1:4]$, $[2:5]$, $[4:6]$, $[5:7]$, and $[8:10]$; each maximal clique is a complete subgraph.}
    \label{fig:max_cliques_decomp}
\end{figure}
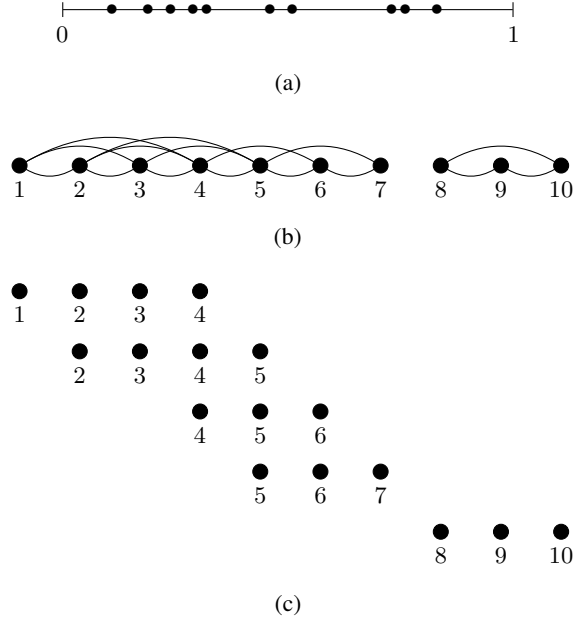

Thus, a maximal clique of $\hat G_{n,r}$ can be identified by its end-nodes only (i.e., those with lowest and highest indices), so that we denote a maximal clique $\{a,a+1,\ldots,b\}$, $a\leq b$, by $[a:b]$. In this way, any ordered graph having $k$ maximal cliques ($1\leq k\leq n$) can be represented by the set $\{[a_1:b_1],\ldots,[a_k:b_k]\}$, where the indices of the end-nodes of the maximal cliques need to satisfy
\begin{equation}\label{eq:cond_max_cliques}
\begin{split}
    & 1=a_1<a_2<\ldots<a_k\leq n, \\
    & 1\leq b_1<b_2<\ldots<b_k=n, \\
    & a_{i+1}\leq b_i+1, \; i=1,\ldots,k-1.
\end{split}
\end{equation}
The above conditions ensure that the maximal cliques cannot include one another, although they may be overlapping, and each vertex belongs to at least one maximal clique. When $a_{i+1}\leq b_i$, the $i$th and $(i+1)$th maximal cliques are overlapping, whereas when $a_{i+1}= b_i+1$ there is a break in the ordered graph between vertices $b_i$ and $b_i+1$. Also, it is possible that $a_i=b_i$ for some $i$, which occurs when node $a_i$ is isolated and thus constitutes, itself, the $i$th maximal clique. The example in Fig.~\ref{fig:max_cliques_decomp} illustrates the decomposition of an ordered graph into its constituent maximal cliques.

\subsection{Ordered Graphs and Structures} \label{sec:og-s}
\begin{lemma}\label{lemma:OG_iso}
If two different, connected ordered graphs in $\hat{\mathcal{G}}_{n,r}$ are isomorphic, then the ``backward identity'' $\pi(i)=n-i+1$, for all $i\in V$, must be an isomorphism between the two ordered graphs.
\end{lemma}
\begin{IEEEproof}
Let $G$ and $G'$ be two connected ordered graphs in $\hat G_{n,r}$, and assume they are different ($G\neq G'$) and isomorphic ($G\simeq G'$). We consider their maximal-clique representations $G \equiv \{[a_1:b_1],\ldots,[a_k:b_k]\}$ and $G' \equiv \{[a'_1:b'_1],\ldots,[a'_{k'}:b'_{k'}]\}$, where the end-nodes satisfy the conditions~\eqref{eq:cond_max_cliques} with the stronger restriction that the strict inequalities $a_{i+1}<b_i+1$ and $a'_{j+1}<b'_j+1$ must hold, because  the graphs are connected and therefore consecutive maximal cliques must overlap. Note that any isomorphism maps each maximal clique of $G$ onto a unique maximal clique of $G'$. Therefore, the two graphs have the same number of maximal cliques, i.e., $k'=k$. Moreover, since consecutive maximal cliques are overlapping, any two successive maximal cliques of $G$ are mapped by any isomorphism onto two consecutive maximal cliques of $G'$. In particular, noting that the first (i.e., leftmost) maximal clique $[1:b_1]$ of $G$ does not have a preceding maximal clique, it follows that $[1:b_1]$  is mapped onto either $[1:b_1']$, in which case $b_1'=b_1$ and there exists an isomorphism $\pi$ that satisfies $\pi(i)=i$, $i=1,\ldots,b_1$, or $[a_k',n]$, in which case $a_k' = n - b_1 + 1$ and there exists an isomorphism $\pi$ that satisfies $\pi(i) = n - i +1$, $i=1,\ldots,b_1$. Then, by considering all the maximal cliques in succession and that $G\neq G'$ by assumption, it results that the backward identity is an isomorphism between the two graphs.
\end{IEEEproof}

\begin{proposition}\label{prop:struct_rep}
The connected graphs in $\mathcal{G}_{n,r}$ that have the same structure have corresponding ordered graphs that are identical up to left-right reversal. 
\end{proposition}
\begin{IEEEproof}
The result follows from Lemma~\ref{lemma:OG_G_iso} and Lemma~\ref{lemma:OG_iso}, together with the fact that composition of isomorphisms is an isomorphism.
\end{IEEEproof}

Similarly to the surjection $f$ introduced earlier, which maps each graph of $\mathcal{G}_{n,r}$ to its corresponding structure in $\mathcal{S}_{n,r}$, we now define the surjective map $\hat{f}:\hat{\mathcal{G}}_{n,r} \to \mathcal{S}_{n,r}$, which associates each ordered graph with its structure. 

Proposition~\ref{prop:struct_rep} enables us to study connected ordered graphs of $\hat{\mathcal{G}}_{n,r}$ as representatives of the connected structures in $\mathcal{S}_{n,r}$, with the understanding that an ordered graph and its image under left-right reversal represent the same structure. Accordingly, for each $S \in \mathcal{S}_{n,r}$ that is connected, it holds that either $|\hat{f}^{-1}(S)| = 1$, when the ordered graph in $\hat{f}^{-1}(S)$ is left-right symmetrical, or $|\hat{f}^{-1}(S)| = 2$, otherwise. 

Disconnected structures can be treated similarly. In this case, permutations of connected components of an ordered graph should also be considered identical, in addition to left-right reversal. Thus, in general, the number of ordered graphs that have a common structure $S$ with $N_S$ connected components satisfies
\begin{equation}
    1 \leq |\hat{f}^{-1}(S)| \leq 2^{N_S} N_S!
\end{equation}

\section{Counting Structures}\label{sec:count}
In this section, we assess the number of distinct structures of the graphs in $\mathcal{G}_{n,r}$, i.e., the cardinality of $\mathcal{S}_{n,r}$, for any number of nodes $n$, connection range $r>0$, and probability distribution of the node locations. Specifically, we exploit the connection between structure and ordered graphs made in the previous section to find an upper bound on $|\mathcal{S}_{n,r}|$.

\subsection{Upper Bounds}
Based on the surjection $f$ defined in Sec.~\ref{sec:prelims}, the number of structures can be expressed as
\begin{equation}
    |\mathcal{S}_{n,r}| = \sum_{G\in\mathcal{G}_{n,r}} \frac{1}{|f^{-1}(f(G))|},
\end{equation}
where the denominator gives the size of the equivalence class the graph $G\in\mathcal{G}_{n,r}$ belongs to. We will however consider the following alternative expression based on the ordered graphs
\begin{equation} \label{eq:UB_S_G}
    |\mathcal{S}_{n,r}| 
    = \sum_{G \in \hat{\mathcal{G}}_{n,r}} \frac{1}{|\hat{f}^{-1}(\hat{f}(G))|}  < |\hat{\mathcal{G}}_{n,r}|, 
\end{equation}
because we are able to find an upper limit for the number of ordered graphs. Moreover, the set $\hat{\mathcal{G}}_{n,r}$ is smaller than $\mathcal{G}_{n,r}$. Specifically, we count the number of objects defined through the conditions~\eqref{eq:cond_max_cliques}, which gives an upper bound on $|\hat{\mathcal{G}}_{n,r}|$ because every ordered graph must satisfy~\eqref{eq:cond_max_cliques}, although the opposite may not necessarily be true (see the discussion in the next subsection). By doing so, we obtain the following upper bound on the number of structures.

\begin{proposition} \label{prop:card_S}
For any number of nodes $n$, spatial distribution of the nodes, and connection range $r>0$, the number of structures in $\mathcal{S}_{n,r}$ is upper bounded by the $n$th Catalan number, i.e.,
\begin{equation}\label{eq:nS}
    |\mathcal{S}_{n,r}| < c_n =\frac{1}{n+1} {2n \choose n} < \frac{4^n}{n^{3/2}\sqrt{\pi}}.
\end{equation}
\end{proposition}
\begin{IEEEproof}
In Appendix~\ref{app:nog}, we show that $|\hat{\mathcal{G}}_{n,r}|$ is upper bounded by the $n$th Catalan number. The first inequality holds because $|\mathcal{S}_{n,r}| < |\hat{\mathcal{G}}_{n,r}|$, as given in~\eqref{eq:UB_S_G}. The final strict inequality follows from the upper bound on the central binomial coefficient ${2n \choose n} < \frac{4^n}{\sqrt{\pi n}}$ (for a proof, see, e.g.,~\cite{Elkies2013}).
\end{IEEEproof}
Regarding the rightmost inequality in~\eqref{eq:nS}, it is well known that the Catalan number $c_n$ and $\frac{4^n}{n^{3/2}\sqrt{\pi}}$ are asymptotically equivalent.

Considering now only connected graphs, let $\hat{\mathcal{G}}_{n,r}^\text{c}$ and $\mathcal{S}_{n,r}^\text{c}$ be the subsets that include all the connected graphs/structures of $\hat{\mathcal{G}}_{n,r}$ and $\mathcal{S}_{n,r}$, respectively. 
The following bounds on the number of connected structures follow from the fact that $\hat{f}^{-1}(S)$ has at most two elements for connected $S$:
\begin{equation}\label{eq:UB_S_G_co}
    \frac{1}{2} |\hat{\mathcal{G}}_{n,r}^\text{c}| < |\mathcal{S}_{n,r}^\text{c}| 
    <  |\hat{\mathcal{G}}_{n,r}^\text{c}|.
\end{equation}

By finding an upper limit for $|\hat{\mathcal{G}}_{n,r}^\text{c}|$, we obtain the following upper bound for $|\mathcal{S}_{n,r}^\text{c}|$. 

\begin{proposition}\label{prop:card_Sc}
For any number of nodes $n$, spatial distribution of the nodes, and connection range $r>0$, the number of connected structures in $\mathcal{S}_{n,r}$ is upper bounded by 
\begin{equation}
    |\mathcal{S}_{n,r}^\text{c}| < c_{n-1} = \frac{1}{n} { 2n-2 \choose n-1 }.
\end{equation}
\end{proposition}
\begin{IEEEproof}
In Appendix~\ref{app:ncog}, we show that $|\hat{\mathcal{G}}_{n,r}^\text{c}| \leq c_{n-1}$. The result follows from~\eqref{eq:UB_S_G_co}.
\end{IEEEproof}

As suggested by one of the reviewers, Propositions~\ref{prop:card_S} and~\ref{prop:card_Sc} indicate that there might be a correspondence between structures and a certain combinatorial interpretation of the Catalan numbers. In Appendix~\ref{app:Dyck}, we provide a combinatorial proof of the results of Propositions~\ref{prop:card_S} and~\ref{prop:card_Sc} by making a link between the objects defined by~\eqref{eq:cond_max_cliques} and Dyck paths~\cite{Deutsch1999}. This complements the algebraic proofs of Appendices~\ref{app:nog} and~\ref{app:ncog}. The combinatorial interpretation in Appendix~\ref{app:Dyck} gives additional results on the structures of the graphs in $\mathcal{G}_{n,r}$, which we summarize here.

\begin{lemma}
The number of structures in $\mathcal{S}_{n,r}$ with exactly $k$ maximal cliques is upper bounded by the Narayana number $N(n,k)= \frac{1}{n} {n \choose k} {n \choose k-1}$.
\end{lemma}

\begin{lemma}
The number of structures in $\mathcal{S}_{n,r}$ with exactly $k$ connected components is upper bounded by $\frac{k}{n} {2n-k-1 \choose n-1}$.
\end{lemma}

\subsection{Improved Bounds}
The upper bounds obtained in the previous subsection do not depend on the connection range $r$. However, the number of ordered graphs and structures that are possible in the studied model depends on $r$ in general. The dependency on $r$ is because the value of $r$ restricts the set of possible maximal cliques, since gaps of at least $r$ must exist between some of the nodes belonging to different maximal cliques. More specifically, if  $[a:b]$ is a maximal clique in the ordered graph, then the separation between nodes $a$ and $b+1$ (and between $a-1$ and $b$) must be of at least $r$. Thus, a given maximal-clique decomposition is realisable only for a certain minimum spread of the points (distance between the first and the last nodes). This implies that there exist objects defined through condition~\eqref{eq:cond_max_cliques} that are not realisable for some values of $r$, i.e., they do not correspond to the maximal-clique decomposition of any graph in $\hat{\mathcal{G}}_{n,r}$. For example, assume a $10$-node  graph with maximal-cliques $[1:3]$, $[2:4]$, $[4:6]$, and $[7:10]$. Since the distances between nodes $1$ and $4$ and between $4$ and $7$ must be greater than $r$, the spread of the underlying points must be of at least $2r$, and therefore the graph is not realisable for $r>0.5$. 

At one extreme is the empty ordered graph, when all nodes are isolated and therefore the graph has to be ``longer'' than $(n-1)r$. Thus, we can say that for $r<\frac{1}{n-1}$ every object defined by~\eqref{eq:cond_max_cliques} corresponds to a maximal-clique decomposition of an order graph, and therefore the number of possible ordered graphs is exactly the $n$th Catalan number, $c_n$, for every $r$ in $(0,\frac{1}{n-1})$.

Next, we find the exact number of ordered graphs for any value of $r$, which gives the following improved bound.

\begin{proposition}\label{prop:nog_exact}
For any number of nodes $n$, spatial distribution of the nodes, and connection range $r>0$, we have 
\begin{equation}
    |\mathcal{S}_{n,r}| < |\hat{\mathcal{G}}_{n,r}| = A_n(\lceil 1/r \rceil),
\end{equation}
where $A_n(h)$, $h\geq 1$, denotes the number of Dyck paths of length $2n$ and height less than or equal to $h$  (or, equivalently, the number of rooted ordered trees on $n+1$ nodes of depth $\leq h+1$), which is given by~\cite{deBrujin1972}
\begin{equation}\label{eq:A}
    A_n(h) = \frac{2^{2n+2}}{h+2} \sum_{1\leq i \leq (h+1)/2} \sin^2{\left(\frac{\pi i}{h+2}\right)} \cos^{2n}{\left(\frac{\pi i}{h+2}\right)}.
\end{equation}
\end{proposition}
\begin{IEEEproof}
The proof is included in Appendix~\ref{app:nog_exact}.
\end{IEEEproof}

When $h \geq n$, $A_n(h)=c_n$ (the $n$th Catalan number), which supports our earlier observation that $|\hat{\mathcal{G}}_{n,r}| = c_n$ when $r \in (0,\frac{1}{n-1})$. Furthermore, $|\hat{\mathcal{G}}_{n,r}|$ as a function of $r$ is constant when $r \in [\frac{1}{k},\frac{1}{k-1})$, for every integer $k \geq 2$, and exhibits jumps at $r=\frac{1}{k}$, $k \geq 1$, as illustrated by Fig.~\ref{fig:UBfinite}. For example, when $r \in [\frac{1}{2},1)$, the number of ordered graphs is a power of two, as~\eqref{eq:A} gives $|\hat{\mathcal{G}}_{n,r}| = 2^{n-1}$; for $r \in [\frac{1}{3},\frac{1}{2})$, the size of $\hat{\mathcal{G}}_{n,r}$ is given by the $(2n-1)$th Fibonacci number, $|\hat{\mathcal{G}}_{n,r}| = F_{2n-1}$; whereas for $r \in [\frac{1}{4},\frac{1}{3})$, we obtain  $|\hat{\mathcal{G}}_{n,r}| = \frac{1}{2} \left( 3^{n-1}+1 \right)$. 

\begin{proposition}\label{prop:ncs_LB}
For any number of nodes $n$, spatial distribution of the nodes, and connection range $r>0$, the number of structures and the number of connected structures are lower bounded by 
\begin{equation}\label{eq:ncs_LB}
    |\mathcal{S}_{n,r}|  > |\mathcal{S}_{n,r}^\text{c}|  > \frac{1}{2} \sum_{k=1}^{\lceil 1/r \rceil} A_{n-k}(k),
\end{equation}
where $A_n(h)$ is given by~\eqref{eq:A}.
\end{proposition}
\begin{IEEEproof}
The proof follows from~\eqref{eq:UB_S_G_co} and the lower bound on $ |\hat{\mathcal{G}}_{n,r}^\text{c}|$ obtained in Appendix~\ref{app:ncog_LB}.
\end{IEEEproof}

The previously obtained bounds give the following asymptotic result.
\begin{proposition}\label{prop:ns_theta}
For any distribution of the points and fixed connection range $r$, both the number of structures and the number of connected structures are $\Theta\left(a^{2n}\right)$ with $a = a(r) = 2 \cos{\left(\frac{\pi}{\lceil 1/r \rceil+2}\right)}$. 
\end{proposition}
\begin{IEEEproof}
For large $n$ and fixed $h$, the term in~\eqref{eq:A} corresponding to $i=1$ dominates because it gives the largest value of the cosine, such that~\eqref{eq:A} simplifies to the following asymptotic formula~\cite{deBrujin1972}
\begin{equation}
    A_n(h) \sim \frac{2^{2n+2}}{h+2} \sin^2{\left(\frac{\pi}{h+2}\right)} \cos^{2n}{\left(\frac{\pi}{h+2}\right)}.
\end{equation}
Similarly, the lower bound~\eqref{eq:ncs_LB} is dominated by the term corresponding to $k=\lceil 1/r \rceil$. Thus, for a fixed connection range $r$ and sufficiently large $n$, there exist constants $k_1(r)$ and $k_2(r)$ such that
\begin{equation}
    k_1(r) \cdot \left[2 \cos{\left(\frac{\pi}{\lceil 1/r \rceil+2}\right)} \right]^{2n} < |\mathcal{S}_{n,r}^\text{c}| < |\mathcal{S}_{n,r}| < k_2(r) \cdot \left[2 \cos{\left(\frac{\pi}{\lceil 1/r \rceil+2}\right)} \right]^{2n}
\end{equation}

\end{IEEEproof}

\section{Bounds on the Structural Entropy}\label{sec:struct_ent}
In the previous section, we characterized the structural complexity of the considered random geometric graph model by studying the number of possible structures in $\mathcal{S}_{n,r}$. That is, all the different structures were equally accounted for  without taking into account the distribution of the node locations. Still, the bounds on the number of structures obtained in the previous section immediately give the following result.
\begin{theorem}\label{th:UBcount}
For any finite number of points $n$, spatial distribution of the points, and connection range $r>0$, the structural entropy is upper bounded by
\begin{equation}\label{eq:UB_finite}
    H(S_{n,r}) < \log_2 A_n(\lceil 1/r \rceil) < 2n - \frac{3}{2} \log_2 n - \frac{1}{2} \log_2 \pi,
\end{equation}
where $A_n$ is given by~\eqref{eq:A}. For large $n$ and fixed $r$,
\begin{equation}
    H(S_{n,r}) \leq 2n \left[ 1 + \log_2 \cos{\left(\frac{\pi}{\lceil 1/r \rceil+2}\right)} \right] + O(1).
\end{equation}
\end{theorem}
\begin{IEEEproof}
The entropy of any distribution is upper bounded by the cardinality of its support, such that $H(\mathcal{S}_{n,r}) \leq \log_2 |\mathcal{S}_{n,r}|$. The result for finite $n$ follows from Proposition~\ref{prop:nog_exact} and Proposition~\ref{prop:card_S}. The asymptotic result follows from Proposition~\ref{prop:ns_theta}.
\end{IEEEproof}

\begin{figure}
    \centering
    \includegraphics[width=0.6\columnwidth]{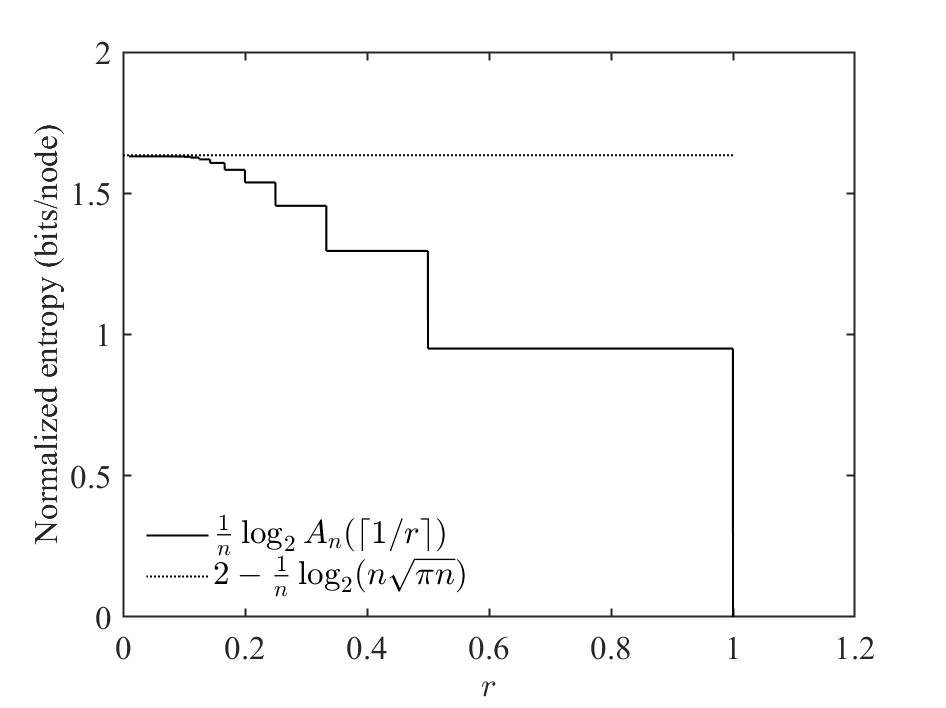}
    \caption{Upper bounds on the normalized structural entropy, $\frac{1}{n}H(S_{n,r})$, given in Theorem~\ref{th:UBcount}, for $n=20$ nodes.}
    \label{fig:UBfinite}
\end{figure}

The bounds for finite $n$ given in~\eqref{eq:UB_finite} are illustrated in Fig.~\ref{fig:UBfinite}. The asymptotic upper bound in Theorem~\ref{th:UBcount} (plotted in Fig.\ref{fig:bounds_fixed_r}) is attained for uniform distribution of the structures. In general, the connection range and the distribution of the node locations dictate how prevalent the different structures are in the ensemble of graphs. For i.u.d. points, the structural entropy approaches zero as $r\to 0$ or $r \to 1$, such that we would expect that, as $r$ increases from zero, the entropy first increases,  reaches a maximum, and then decreases towards zero. However, the upper bound given by Theorem~\ref{th:UBcount} is nonincreasing with $r$. In the following, we focus on the structural entropy $H(S_{n,r})$ given by~\eqref{eq:struct_entropy_def} and thus explicitly consider the distribution of the locations. 

\subsection{Relationship Between $H(S_{n,r})$ and $H(\hat{G}_{n,r})$}
We further exploit the connection between the random structure $S_{n,r}$ and the ordered graph $\hat{G}_{n,r}$ to obtain upper and lower bounds on $H(S_{n,r})$.

\begin{lemma}\label{lemma:UB}
For any number of nodes $n$, distribution of the locations, and connection range $r$, 
\begin{equation}\label{eq:HsHg}
    H(S_{n,r}) \leq H(\hat{G}_{n,r}).
\end{equation}
\end{lemma}
\begin{IEEEproof}
According to the map $\hat{f}$ introduced in Sec.~\ref{sec:og-s},
\begin{equation}
    H(S_{n,r}) = H ( \hat{f}(\hat{G}_{n,r}) ) \leq H ( \hat{G}_{n,r} ),
\end{equation}
where the inequality holds because the entropy of deterministic functions of a random variable is no larger than the entropy of the variable~\cite{Cover2006}.
\end{IEEEproof}

When considering only the connected graphs, we find an upper bound and a lower bound that determine the structural entropy within one bit. 
\begin{lemma}\label{lemma:LUBc}
Let $S_{n,r}^\text{c}$ denote the random structure conditioned on it being connected. The connected ordered graph $\hat{G}_{n,r}^\text{c}$ is similarly defined. Then, for any number of nodes $n$, distribution of the locations, and connection range $r$,
\begin{equation}
    H(\hat{G}_{n,r}^\text{c}) - 1  \leq H(S_{n,r}^\text{c}) \leq  H(\hat{G}_{n,r}^\text{c}).
\end{equation}
\end{lemma}
\begin{IEEEproof}
As established earlier, the preimage $\hat{f}^{-1}(S)$ for connected structures contains only one element when $S$ is left-right symmetrical, and two elements, otherwise. Thus, for any $S\in\mathcal{S}_{n,r}^\text{c}$, \eqref{eq:Ps} gives
\begin{equation}\label{eq:Psg}
    P(S) = 
    \begin{cases}
    P(\hat{G}), &\text{if } S \text{ is symm.}, S = \hat{f}(\hat{G}), \\
    P(\hat{G}_1)+P(\hat{G}_2), &\text{if } S \text{ is asymm.}, \{\hat{G}_1, \hat{G}_2\} \equiv \hat{f}^{-1}(S).
    \end{cases}
\end{equation}
We now write
\begin{align} \label{eq:HGS}
    H(\hat{G}_{n,r}^\text{c})
    & = -\sum_{\hat{G} \in \hat{\mathcal{G}}_{n,r}^\text{c}} P(\hat{G}) \log_2 P(\hat{G}) \nonumber \\
    &=  -\sum_{S \in \mathcal{S}_{n,r}^\text{c}} \sum_{\hat{G} \in f^{-1}(S)} P(\hat{G}) \log_2 P(\hat{G}).
\end{align}
For any asymmetrical structure $S$, define $q_S = \frac{1}{P(S)}\min(P(\hat{G}_1), P(\hat{G}_2))$, where $\{\hat{G}_1, \hat{G}_2\} \equiv \hat{f}^{-1}(S)$. From~\eqref{eq:Psg}, $q_S < 1$. Using~\eqref{eq:Psg}, we further express~\eqref{eq:HGS} as
\begin{align}
    H(\hat{G}_{n,r}^\text{c})
    & =  -\sum_{S \in \mathcal{S}_{n,r}^\text{c}} P(S) \log_2 P(S)    -\sum_{\substack{S \in \mathcal{S}_{n,r}^\text{c} \\ S \text{ asymm.}}} P(S) \left[ q_S \log_2 q_S + (1-q_S) \log_2(1-q_S) \right] \nonumber \\
    &= H(S_{n,r}^\text{c}) + \sum_{\substack{S \in \mathcal{S}_{n,r}^\text{c} \\ S \text{ asymm.}}} P(S) H_\text{b}(q_S),
\end{align}
where $H_\text{b}(\cdot)$ is the binary entropy function. Given that $H_\text{b}$ takes on values smaller than one and $P(S)$ is a probability distribution, the sum term above is between zero and one, and the desired result follows.
\end{IEEEproof}

Lemmas~\ref{lemma:UB} and~\ref{lemma:LUBc} give upper and lower bounds on the structural entropy in terms of the entropy of the ordered graph, $H(\hat{G}_{n,r})$. In the following, we find bounds on $H(\hat{G}_{n,r})$ and study them for i.u.d. point locations and different scaling functions $r_n$, results which ultimately transfer to bounds for the structural entropy.

\subsection{Bounds on $H(\hat{G}_{n,r})$}\label{sec:UB}
\begin{figure}
    \centering
        \def\d{0.8}
        \begin{tikzpicture}
            \footnotesize
            \node[] (N1)   at (0*\d,0) {...};
            \node[fill,circle,inner sep=2pt,outer sep=0pt,draw,label=below:$i-L_i$] (N2)   at (1*\d,0) {};
            \node[] (N3)   at (2*\d,0) {...};
            \node[fill,circle,inner sep=2pt,outer sep=0pt,draw,label=below:$i+1-L_{i+1}$] (N4)   at (3*\d,0) {};
            \node[] (N5)   at (4*\d,0) {...};
            \node[fill,circle,inner sep=2pt,outer sep=0pt,draw,label=below:$i$] (N6)   at (5*\d,0) {};
            \node[fill,circle,inner sep=2pt,outer sep=0pt,draw,label=below:$i+1$] (N7)   at (6*\d,0) {};
            \node[] (N8)   at (7*\d,0) {...};
            \draw[bend left] (N2) to node[]{} (N6);
            \draw[bend left] (N4) to node[]{} (N6);
            \draw[bend left] (N4) to node[]{} (N7);
            \draw[bend left] (N6) to node[]{} (N7);
    \end{tikzpicture}
    \caption{In this illustration, node $i-L_i$ is the leftmost neighbour of node $i$, i.e., node $i$ is connected by an edge to nodes $i-L_i,i-L_i+1,\ldots,i-1$ but not to nodes $1,\ldots,i-L_i-1$. Similarly, node $i+1$ has $L_{i+1}$ neighbours to the left.}
    \label{fig:Li}
\end{figure}
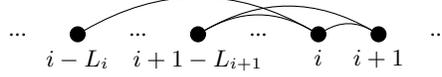

We represent $\hat{G}_{n,r}$ in terms of the number of leftward neighbours the nodes have, as follows. Recall that the vertices of $\hat{G}_{n,r}$ are indexed in increasing order from left to right, according to the order of their locations. For all $i \in \{2,\ldots,n\}$, let $L_i$ be the number of neighbours that node $i$ has to the left, which satisfies $0 \leq L_i \leq i-1$. It can be inspected (e.g., see Fig.~\ref{fig:Li}) that  $0 \leq L_{i+1} \leq L_i + 1$.

As there is a one-to-one relationship between $\hat{G}_{n,r}$ and $(L_2,\ldots,L_n)$, and according to the chain rule for entropy, we have
\begin{align}\label{eq:HLchain}
    H(\hat{G}_{n,r}) &= H(L_2,\ldots,L_n) \nonumber \\
    &= H(L_2) + \sum_{i=2}^{n-1} H(L_{i+1} \mid L_2,\ldots, L_i),
\end{align}
\begin{lemma}
The following upper bound holds for any number $n$ of points, spatial distribution of the points, and connection range $r$:
\begin{equation} \label{eq:H_UB_chain}
    H(\hat{G}_{n,r}) \leq H(L_2) + \sum_{i=2}^{n-1} H(L_{i+1} \mid L_i).
\end{equation}
\end{lemma}
\begin{IEEEproof}
Given that conditioning reduces the entropy, the result immediately follows from~\eqref{eq:HLchain}.
\end{IEEEproof}
\begin{lemma}
When the point locations $X_1,\ldots,X_n$ are independently sampled from a continuous distribution, the following lower bound holds for any number $n$ of points and connection range $r$:
\begin{equation} \label{eq:H_LB_chain}
    H(\hat{G}_{n,r}) \geq H(L_2) + \sum_{i=2}^{n-1} H\left(L_{i+1} \mid L_i, X_{(i-L_i)},\ldots,X_{(i)} \right).
\end{equation}
\end{lemma}
\begin{IEEEproof}
Since conditioning reduces the entropy, the terms of the sum in~\eqref{eq:HLchain} are lower bounded by
\begin{align}
    H(L_{i+1} \mid L_2,\ldots,L_i) 
    &\geq H\left(L_{i+1} \mid L_2,\ldots,L_i, X_{(i-L_i)},\ldots,X_{(i)} \right) \\
    &= H\left(L_{i+1} \mid L_i, X_{(i-L_i)},\ldots,X_{(i)} \right)
\end{align}
The equality holds because, given $L_i$, node $i+1$ may be connected by an edge only to nodes $i-L_i,i-L_i+1,\ldots,i$ and thus $L_{i+1}$ is dictated by the gap $X_{(i+1)} - X_{(i)}$ and the locations $X_{(i-L_i)},\ldots,X_{(i)}$. Then, given that the order statistics of independent variables drawn from a continuous distribution have the Markov property~\cite{David1981}, the gap $X_{(i+1)} - X_{(i)}$ is conditionally independent of $L_2,\ldots,L_i$, given $X_{(i)}=x$;  consequently, $L_{i+1}$ is conditionally independent of $(L_2,\ldots,L_{i-1})$, given the number of leftward neighbors $L_i$ and the locations $X_{(i-L_i)},\ldots,X_{(i)}$.
\end{IEEEproof}

Next, we evaluate the obtained bounds for i.u.d. point locations and different scaling functions $r_n$. 

\begin{proposition}\label{prop:UBunif}
When the point locations $X_1,\ldots,X_n$ are i.u.d. on $[0,1]$, the upper bound on the normalized entropy of the ordered graph satisfies the asymptotic equivalence
\begin{align}\label{eq:UBunif}
    \frac{1}{n} H(\hat{G}_{n,r_n}) 
     &\leq \frac{1}{n} H(L_2) + \frac{1}{n} \sum_{i=2}^{n-1} H(L_{i+1} \mid L_i) \nonumber \\
    &\sim 
    \begin{cases}
    h_\mathrm{U}(n r_n), &\text{if } r_n = O(1/n),\\
    2, & \text{if } r_n = \omega(1/n) \text{ and } r_n = o(1), \\
    2(1-r), & \text{if } r_n = r \in (0,1),
    \end{cases}
\end{align}
as $n \to \infty$, where the function $h_\mathrm{U}:[0,\infty) \to \mathbb{R}_+$ is given by
\begin{equation}\label{eq:hU_func}
    h_\mathrm{U}(x) = \frac{x \, e^{-x}}{\ln 2}  - \sum_{a=0}^{\infty} \frac{x^a e^{-x}}{a!} \sum_{k=0}^{a} \frac{x}{a+1} M(k+1,a+2,-x) \log_2\left(\frac{x}{a+1} M(k+1,a+2,-x)\right)
\end{equation}
with $M$ being Kummer's confluent hypergeometric function,
\begin{equation*}
    M(\alpha,\beta,z) = \frac{\Gamma(\beta)}{\Gamma(\alpha) \, \Gamma(\beta-\alpha)} \int_0^1 t^{\alpha-1} (1-t)^{\beta-\alpha-1} e^{zt} \mathrm{d}t.
\end{equation*}
\end{proposition}
\begin{IEEEproof}
See Appendix~\ref{app:proof_UBunif}.
\end{IEEEproof}

\begin{proposition}\label{prop:LBunif}
When the point locations $X_1,\ldots,X_n$ are i.u.d. on $[0,1]$, the lower bound on the normalized entropy of the ordered graph satisfies the asymptotic equivalence
\begin{align}\label{eq:LBunif}
     \frac{1}{n} H(\hat{G}_{n,r_n}) 
     &\geq \frac{1}{n} H(L_2) + \frac{1}{n} \sum_{i=2}^{n-1} H\left(L_{i+1} \mid L_i, X_{(i-L_i)},\ldots,X_{(i)} \right)  \nonumber \\
    &\sim 
    \begin{cases}
    h_\mathrm{L}(n r_n), &\text{if } r_n = O(1/n),\\
    \log_2 e, & \text{if } r_n = \omega(1/n) \text{ and } r_n = o(1),\\
    (1-r)\log_2 e, & \text{if } r_n = r \in (0,1),
    \end{cases}
\end{align}
as $n \to \infty$, where the function $h_\mathrm{L}:[0,\infty) \to \mathbb{R}_+$ is given by
\begin{equation}
    h_\mathrm{L}(x) = \left( 1-e^{-x} \right) \log_2 e  - \left( 1-e^{-x} \right)^2 \log_2 \left( 1-e^{-x} \right).
\end{equation}
\end{proposition}
\begin{IEEEproof}
See Appendix~\ref{app:proof_LBunif}.
\end{IEEEproof}

Fig.~\ref{fig:bounds} illustrates the upper and lower bounds given by Propositions~\ref{prop:UBunif} and~\ref{prop:LBunif} when the connection range function is $r_n = \frac{x}{n}$, $x>0$, and $n$ is large. We notice that the bounds approach the values of $2$ bits and respectively $\log_2 e \approx 1.443$ bits, as $x$ increases; these values correspond to the asymptotic value (in $n$) of the normalized entropy of the ordered graph when $r_n$ dominates $1/n$ but is still vanishing with $n$ (e.g., $r_n \propto \ln n / n$ or $r_n \propto 1/\sqrt{n}$). Finally, if the scaling function is constant, $r_n=r$ with $r$ bounded away from $0$ and $1$, the bounds given by Propositions~\ref{prop:UBunif} and~\ref{prop:LBunif} decrease linearly with $r$. The decrease is due to the margin effect caused by the finite domain over which the points are distributed: roughly speaking, for large $n$, approximately a fraction $r$ of the nodes form the leftmost clique (i.e., they are all connected to node $1$), such that only a fraction of about $1-r$ of the $L$ variables contribute to the uncertainty. 

\begin{figure}
    \centering
    \includegraphics[width=0.6\columnwidth]{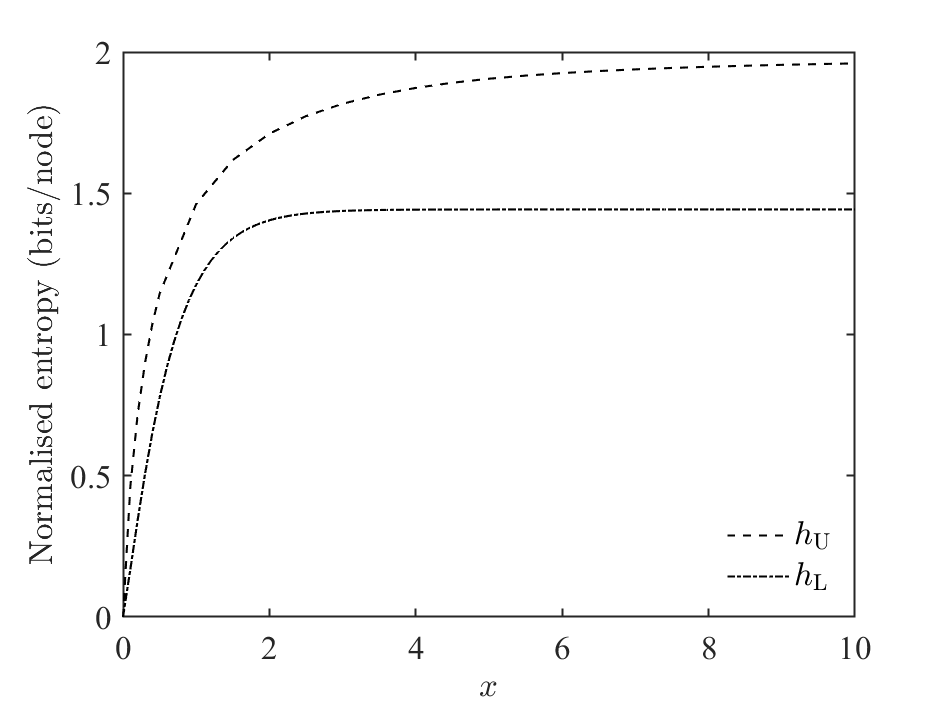}
    \caption{Upper and lower bounds on the normalized entropy of the ordered graph given by Propositions~\ref{prop:UBunif} and~\ref{prop:LBunif} when $r_n = \frac{x}{n}$ and $n\to\infty$. }
    \label{fig:bounds}
\end{figure}

\subsection{Structural Entropy for i.u.d. Points}
The results obtained in this section are summarized in terms of the structural entropy in the following theorem.
\begin{theorem}\label{th:iud}
When the locations $X_1,\ldots,X_n$ are i.u.d. over $[0,1]$, the normalized structural entropy is upper bounded by~\eqref{eq:UBunif}, as $n\to \infty$. When the connection range function $r_n$ is vanishing but stays above the critical scaling, such that the graph is connected with probability one (i.e., $r_n$ is in the form required in~\eqref{eq:Pcon}), the normalized structural entropy is bounded by
\begin{equation}
    \log_2 e < \frac{1}{n} H(S_{n,r}) < 2, \quad \text{as } n \to \infty.
\end{equation}
Furthermore, when $r_n=r$ with $r$ a fixed constant in $(0,1)$, the normalized structural entropy is bounded by
\begin{equation}
    (1-r) \log_2 e < \frac{1}{n} H(S_{n,r}) < 2(1-r), \quad \text{as } n \to \infty.
\end{equation}

\end{theorem}
\begin{IEEEproof}
The first part follows from Lemma~\ref{lemma:UB} and Proposition~\ref{prop:UBunif}. When $r_n$ is above the critical threshold, Lemma~\ref{lemma:LUBc} gives that the normalized structural entropy is equal to the normalized entropy of the ordered graph as $n\to\infty$. Then, the upper and lower limits follow from Propositions~\ref{prop:UBunif} and~\ref{prop:LBunif}.
\end{IEEEproof}

\begin{figure}
    \centering
    \includegraphics[width=0.6\columnwidth]{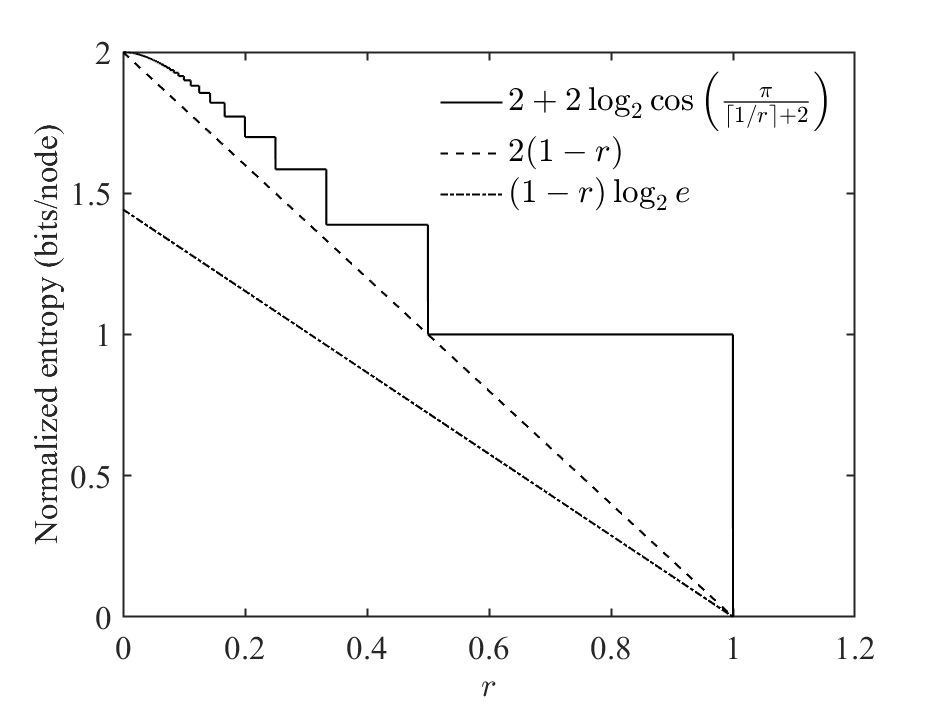}
    \caption{Upper and lower bounds on the normalized structural entropy given by Theorem~\ref{th:UBcount} and Theorem~\ref{th:iud} for fixed $r$ and $n\to\infty$. }
    \label{fig:bounds_fixed_r}
\end{figure}

The asymptotic bounds on the structural entropy given by Theorem~\ref{th:UBcount} and Theorem~\ref{th:iud} for fixed connection range are illustrated in Fig.~\ref{fig:bounds_fixed_r}.

Thus, in the asymptotically connected regime, it should be possible to encode the random structures using no more than $2n$ bits. A simple encoding scheme that is based on the maximal clique representation (see Sec.~\ref{sec:max_cliques}) and requires $2n$ bits is as follows. We use two binary strings $\mathbf{a}$ and $\mathbf{b}$, each of length $n$. The ones give the indices of the end-nodes of the maximal cliques. Specifically, the position of the $k$th one in $\mathbf{a}$ ($\mathbf{b}$) gives the index of the leftmost (rightmost) node of the $k$th maximal clique. For example, the graph in Fig.~\ref{fig:max_cliques_decomp} is encoded as $\mathbf{a} = 1101100100$ and $\mathbf{b} = 0001111001$.

\section{Conclusions}
In this paper, we examined the complexity of the ensemble of structures (unlabeled graphs) of the one-dimensional random geometric graph model. We introduced the ordered graph, which has the nodes labeled according to the order of their underlying locations, and showed that the connected ordered graphs of the model can be used as representatives of the connected structures, up to left-right reversal. A disconnected structure can also be represented by the ordered graph by additionally discounting the possible permutations of the connected components. Based on the decomposition of the ordered graph into its constituent maximal cliques, we found that the number of ordered graphs and thus of possible structures in the 1D RGG model is upper limited by the $n$th Catalan number, which determined us to identify an interesting connection between ordered graphs and known combinatorial objects, such as Dyck paths and rooted ordered trees. This result immediately gave an upper bound on the structural entropy of $2n-\frac{3}{2}\log_2 n - \frac{1}{2}\log_2\pi$ which is universal, in the sense that it holds for any distribution of the point locations and connection range. However, the upper bound does not depend on the connection range $r$, whereas the value of $r$ restricts the number of possible ordered graphs. By counting the exact number of ordered graphs for given $r$, we then obtained that the number of structures is $\Theta\left(a^{2n}\right)$ with $a = a(r)$, and an improved universal upper bound on the number of structures, which for is attained for uniform distribution of the structures when $r$ is fixed.

By establishing relationships between the structural entropy and the entropy of the ordered graph, we then investigated the influence of the connection range on the structural entropy for independent and uniformly distributed points, in the limit of large number of nodes. The obtained upper bound exhibits the expected behaviour of the structural entropy, i.e., as $r$ increases from zero, the structural entropy first increases starting from zero but then decreases when $r$ is bounded away from zero, due to the finite embedding domain. In the asymptotically connected regime, the structural entropy normalized by the number of nodes is between $\log_2 e \approx 1.44$ and $2$ bits per node when $r_n$ is vanishing, and between $(1-r)\log_2 e$ and $2(1-r)$ for fixed $r$. A relevant next step would be to find a precise enough asymptotic expansion for the entropy of the ordered graph that would give a more accurate characterization of the normalized structural entropy (up to $o(1)$). We also proposed a simple encoding scheme that reflects the decomposition into maximal cliques and requires $2n$ bits.

The upper bounds we obtained in this paper translate to the entropy of the labeled graph model by adding $\log_2(n!) = n \log_2 n - n + O(\log_2 n)$ to the structural entropy to account for the number of possible permutation of the structures; thus, the term $ n \log_2 n$ dominates the entropy of the labeled graph. It would be relevant to improve this term by considering the sizes of the automorphism groups of the graphs. 

We intend to extend the analysis to RGGs in two (or higher) dimensions. We think it is also relevant to study RGGs with soft connection functions (i.e., where pair of nodes are connected by an edge with a probability that decreases with the distance between them), as opposed to the hard connection model assumed in this paper. 

\bibliographystyle{IEEEtran}
\bibliography{references.bib}

\appendices

\section{Number of Ordered Graphs} \label{app:nog}
Let $\mathcal{C}_n$ be the set of graphs with vertices $\{1,\ldots,n\}$ that have the property that, if an edge exists between two vertices $i$ and $j$, $i<j$, then the vertices $\{i,i+1,\ldots,j\}$ form a clique, for any $i$ and $j$. According to Lemma~\ref{lemma:OG_clique}, $\hat{\mathcal{G}}_{n,r} \subseteq \mathcal{C}_n$ and therefore
\begin{equation}\label{eq:G_C}
    |\hat{\mathcal{G}}_{n,r}| \leq |\mathcal{C}_n|.
\end{equation}
In the rest of this subsection we determine a closed-form expression for $|\mathcal{C}_n|$, which will be used in Proposition~\ref{prop:card_S}.

We define the set $\mathcal{C}_{n,k}$ to include the graphs of $\mathcal{C}_n$ that have $k$ maximal cliques, $k=1,\ldots,n$. Thus, $|\mathcal{C}_n| = \sum_{k=1}^n |\mathcal{C}_{n,k}|$. Every graph in $\mathcal{C}_{n,k}$ can be represented as the set of $k$ maximal cliques $\{[a_1:b_1],\ldots,[a_k:b_k]\}$, where the indices of the end-nodes must satisfy the conditions given in~\eqref{eq:cond_max_cliques}. To compute $|\mathcal{C}_{n,k}|$, we first determine the number $p(m,n,k)$ of graphs in $\mathcal{C}_{n,k}$ whose $k$th (i.e., last) maximal clique is fixed to $[a_k,b_k]=[m,n]$. When $k=1$, all nodes must belong to the single clique, such that
\begin{equation}\label{eq:p_k1}
    p(m,n,1)=
    \begin{cases}
    1, & m=1, \\
    0, & m>1.
    \end{cases}
\end{equation}
For $k\geq 2$, $m$ must satisfy $k \leq m \leq n$. We obtain the following result.
\begin{lemma}\label{lemma:p}
The number of graphs in $\mathcal{C}_{n,k}$, $k\geq 2$, whose last maximal clique is fixed to $[a_k,b_k]=[m,n]$, $k \leq m \leq n$, is 
\begin{equation}\label{eq:p}
    p(m,n,k) = {n\choose k-1}{m-2 \choose k-2} - {n-1\choose k-2}{m-1 \choose k-1}
\end{equation}
\end{lemma}
\begin{IEEEproof}
The proof is by induction over $k$. From the conditions~\eqref{eq:cond_max_cliques} (and by inspecting the decomposition example in Fig.~\ref{fig:max_cliques_decomp}), we establish the following recurrence relation over the number of maximal cliques:
\begin{equation}\label{eq:rec_p}
    p(m,n,k+1) = \sum_{u=k}^{m-1} \sum_{v=m-1}^{n-1} p(u,v,k).
\end{equation}
Using~\eqref{eq:p_k1} in~\eqref{eq:rec_p}, we obtain
\begin{equation}
    p(m,n,2) = \sum_{u=1}^{m-1} \sum_{v=m-1}^{n-1} p(u,v,1) = \sum_{v=m-1}^{n-1} 1 = n-m+1.
\end{equation}
The same expression is obtained by substituting $k=2$ in~\eqref{eq:p}. Thus, the claimed result~\eqref{eq:p} is satisfied for $k=2$. 

Now, assume~\eqref{eq:p} holds for some $k>2$. Eq.~\eqref{eq:rec_p} becomes
\begin{align}
    p(m,n,k+1) 
    &= \sum_{u=k}^{m-1} {u-2 \choose k-2}  \sum_{v=m-1}^{n-1} {v\choose k-1} \nonumber \\
    &\phantom{=} - \sum_{u=k}^{m-1} {u-1 \choose k-1}  \sum_{v=m-1}^{n-1} {v-1\choose k-2}.\label{eq:p_sums}
\end{align}
We next calculate the sums in~\eqref{eq:p_sums}. Using the identity for the rising sum of binomial coefficients, we obtain
\begin{equation}
    \sum_{u=k}^{m-1} {u-2 \choose k-2} = \sum_{i=0}^{m-k-1} {k-2+i \choose k-2} = {m-2 \choose k-1}
\end{equation}
and, similarly,
\begin{equation}
    \sum_{u=k}^{m-1} {u-1 \choose k-1} = {m-1 \choose k}.
\end{equation}
Then, we write
\begin{align}
    \sum_{v=m-1}^{n-1} {v\choose k-1} 
    &= \sum_{v=k-1}^{n-1} {v\choose k-1} - \sum_{v=k-1}^{m-2} {v\choose k-1} \\
    &= {n\choose k} - {m-1\choose k}
\end{align}
and similarly obtain
\begin{equation}
    \sum_{v=m-1}^{n-1} {v-1\choose k-2} = {n-1\choose k-1} - {m-2\choose k-1}.
\end{equation}
Finally, plugging these results back in~\eqref{eq:p_sums}, we obtain
\begin{equation}
    p(m,n,k+1)={n\choose k}{m-2 \choose k-1} - {n-1\choose k-1}{m-1 \choose k}.
\end{equation}
This establishes the inductive step, and thus the proof is complete.
\end{IEEEproof}
\begin{lemma} \label{lemma:Cnk}
The number of graphs in $\mathcal{C}_{n,k}$, $1 \leq k \leq n$, is 
\begin{equation}\label{eq:Cnk}
    |\mathcal{C}_{n,k}| = \frac{1}{n} {n \choose k}{n \choose k-1}.
\end{equation}
\end{lemma}
\begin{IEEEproof}
For $k=1$, the set $\mathcal{C}_{n,1}$ contains only the complete graph, such that $|\mathcal{C}_{n,1}|=1$. For $k\geq 2$, the $k$th maximal clique can start anywhere between $k$ and $n$, i.e., $k\leq a_k\leq n$. Therefore, we have
\begin{align*}
    |\mathcal{C}_{n,k}| 
    &= \sum_{m=k}^n p(m,n,k) \\
    &=  {n\choose k-1}\sum_{m=k}^n{m-2 \choose k-2} - {n-1\choose k-2}\sum_{m=k}^n{m-1 \choose k-1} \\
    &= {n\choose k-1}{n-1 \choose k-1} - {n-1\choose k-2}{n \choose k}\\
    &= \frac{1}{n} {n\choose k}{n \choose k-1}.
\end{align*}
\end{IEEEproof}
\begin{lemma}\label{lemma:Cn}
The number of graphs in $\mathcal{C}_n$ is given by the $n$th Catalan number,
\begin{equation}
    |\mathcal{C}_n| = c_n = \frac{1}{n+1} { 2n \choose n }.
\end{equation}
\end{lemma}
\begin{IEEEproof}
We have 
\begin{align}\label{eq:hG1}
    |\mathcal{C}_n| 
    &= \sum_{k=1}^n |\mathcal{C}_{n,k}| \\ 
    &= \frac{1}{n} \sum_{k=1}^n {n \choose k}{n \choose k-1} \\ 
    &= \frac{1}{n} \sum_{k=1}^n {n \choose k}{n \choose n+1-k} \\ 
    &= \frac{1}{n} { 2n \choose n+1 } \\
    &= \frac{1}{n+1} { 2n \choose n },
\end{align}
where we have used Vandermonde's identity. 
\end{IEEEproof}

\section{Number of Connected Ordered Graphs} \label{app:ncog}
Starting from the set of graphs $\mathcal{C}_n$ defined in Appendix~\ref{app:nog}, let $\mathcal{C}_n^\text{c}$ be the subset consisting of all the connected graphs of $\mathcal{C}_n$  and let $\mathcal{C}_{n,k}^\text{c}$ include all the graphs of $\mathcal{C}_n^\text{c}$ that have $k$ maximal cliques, $1\leq k \leq n-1$.\footnote{Connected $n$-node graphs cannot have $n$ maximal cliques, because that would correspond to an empty graph. An example of a graph with $k=n-1$ maximal cliques is one where each maximal clique consists of two nodes and every two successive maximal cliques have one node in common.} 

Similarly to before, Lemma~\ref{lemma:OG_clique} gives
\begin{equation}\label{eq:G_C_co}
    |\hat{\mathcal{G}}_{n,r}^\text{c}| \leq |\mathcal{C}_n^\text{c}| = \sum_{k=1}^{n-1} |\mathcal{C}_{n,k}^\text{c}|.
\end{equation}
In the maximal-clique representation of each graph in $\mathcal{C}_{n,k}^\text{c}$, $\{[a_1:b_1],\ldots,[a_k:b_k]\}$, the indices $a$ and $b$ must satisfy the conditions given in~\eqref{eq:cond_max_cliques} with the stronger requirement that $a_{i+1} \leq b_i$, for all $i=1,\ldots,k-1$, because consecutive maximal cliques must be overlapping. We define the number $q(m,n,k)$ of graphs in $\mathcal{C}_{n,k}^\text{c}$ whose $k$th maximal clique is fixed to $[a_k,b_k]=[m,n]$. When $k=1$, $q(1,n,1) = 1$ and $q(m,n,1) = 0$, for $m>1$. For $k\geq 2$, $m$ must satisfy $k \leq m \leq n-1$. The following result is the counterpart of Lemma~\ref{lemma:p}.
\begin{lemma}
The number of graphs in $\mathcal{C}_{n,k}^\text{c}$, $k\geq 2$,  with $[a_k,b_k]=[m,n]$, $k \leq m \leq n-1$, is 
\begin{equation}\label{eq:q}
    q(m,n,k) = {n-1 \choose k-1}{m-2 \choose k-2} - {n-2\choose k-2}{m-1 \choose k-1}
\end{equation}
\end{lemma}
\begin{IEEEproof}
The following recursion is found to hold (similarly to the recursion for $p$ in the proof of Lemma~\ref{lemma:p}):
\begin{equation}
    q(m,n,k+1) = \sum_{u=k}^{m-1} \sum_{v=m}^{n-1} q(u,v,k).
\end{equation}
The proof is then by induction over $k$ and involves similar steps as for Lemma~\ref{lemma:p}, which we omit. 
\end{IEEEproof}

\begin{lemma}
The number of graphs in $\mathcal{C}_{n,k}^\text{c}$, $1\leq k \leq n-1$, is 
\begin{equation}
    |\mathcal{C}_{n,k}^\text{c}| = \frac{1}{n-1} {n-1\choose k}{n-1 \choose k-1}.
\end{equation}
\end{lemma}
\begin{IEEEproof}
We omit the proof, as it is similar to that of Lemma~\ref{lemma:Cnk}.
\end{IEEEproof}

\begin{lemma}
The number of graphs in $\mathcal{C}_n^\text{c}$ is given by the $(n-1)$th Catalan number,
\begin{equation}
    |\mathcal{C}_n^\text{c}| = c_{n-1} = \frac{1}{n} { 2n-2 \choose n-1 }.
\end{equation}
\end{lemma}
\begin{IEEEproof}
The proof is similar to that of Lemma~\ref{lemma:Cn}. 
\end{IEEEproof}

\section{Combinatorial Proof of Propositions~\ref{prop:card_S} and~\ref{prop:card_Sc}} \label{app:Dyck}
A Dyck path of length $2n$ is a path in the plane integer lattice $\mathbb{Z}^2$ from the origin $(0,0)$ to $(2n,0)$ that consists of up steps $u = (1,1)$ and down steps $d = (1,-1)$, and never goes below the horizontal axis. Thus, every Dyck path of length $2n$ consists of $2n$ steps of which exactly $n$ are up steps and $n$ are down steps, and after each step the current number of $u$'s is not less than the current number of $d$'s. 

\begin{figure}
    \centering
    \begin{tikzpicture}[scale=0.75]
        \footnotesize
        \foreach \i in {0,...,20} {
            \draw [very thin] (\i,0) -- (\i,5)  node [below] at (\i,0) {$\i$};
        }
        \foreach \i in {0,...,5} {
            \draw [very thin] (0,\i) -- (20,\i) node [left] at (0,\i) {$\i$};
        }
        \draw [very thick] (0,0) -- (4,4) -- (5,3) -- (6,4) -- (8,2) -- (9,3) -- (10,2) -- (11,3) -- (14,0) -- (17,3) -- (20,0);
    \end{tikzpicture}
    \caption{The Dyck path corresponding to the ordered graph depicted in Fig.~\ref{fig:max_cliques_decomp}. The ordered graph consists of five maximal cliques ($[1:4]$, $[2:5]$, $[4:6]$, $[5:7]$, and $[8:10]$), each giving a peak of height equal to the size of the maximal clique. The return of the path to the horizontal axis after the $14$th step corresponds to the gap between nodes $7$ and $8$ of the graph (the $4$th and $5$th maximal cliques are not overlapping).}
    \label{fig:Dyck_path}
\end{figure}
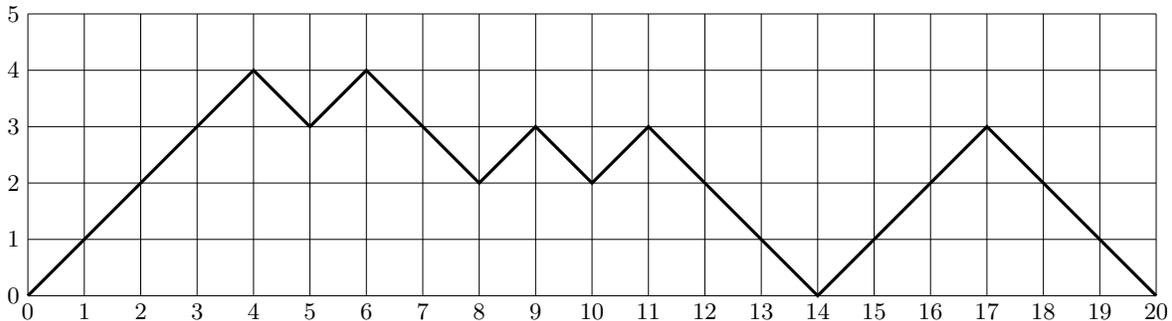

Assume the set of integer intervals $I=\{[a_1:b_1],\ldots,[a_k:b_k]\}$ satisfying conditions~\eqref{eq:cond_max_cliques}, $1\leq k\leq n$. Let us define a map $\varphi$ that maps the set of admissible $I$'s to the set of Dyck paths of length $2n$. We write $u^m = \underbrace{uu\ldots u}_{m\text{ times}}$ for a sequence of $m$ up steps (and similarly for down steps). The map is given by
\begin{equation}\label{eq:map_Dyck}
    \varphi(I) = u^{b_1-a_1+1} d^{a_2-a_1} u^{b_2-b_1}\ldots d^{a_k-a_{k-1}} u^{b_k-b_{k-1}} d^{b_k-a_k+1}.
\end{equation}
The number of up steps in the map definition~\eqref{eq:map_Dyck} is $(b_1-a_1+1) + (b_2-b_1)+\ldots+(b_k-b_{k-1}) = b_k-a_1+1 = n$. Similarly, it can be verified that the number of down steps is also equal to $n$. Furthermore, the height of the $i$th valley (a valley is a $d$ followed by an $u$) of the path described by~\eqref{eq:map_Dyck} is given by the difference  between the number of $u$'s and the number of $d$'s up to the $i$th time the succession $du$ occurs, i.e., $(b_1-a_1+1) + (b_2-b_1) + \ldots + (b_i-b_{i-1}) - (a_2-a_1) - \ldots - (a_{i+1}-a_i) = 1+b_i - a_{i+1}$, which is nonnegative according to condition~\eqref{eq:cond_max_cliques}. Thus, every $\varphi(I)$ is mapped by $\varphi$ on a unique Dyck path of length $2n$. Moreover, it can be inspected that for every Dyck path a unique inverse can be determined from~\eqref{eq:map_Dyck}, such that $\varphi$ is a bijection. In conclusion, the number of objects defined through~\eqref{eq:cond_max_cliques} is equal to the number of Dyck paths of length $2n$, which is well known to be given by $c_n$, the $n$th Catalan number~\cite{Stanley2015}.

As an example, Fig.~\ref{fig:Dyck_path} illustrates the Dyck path corresponding to the ordered graph in Fig.~\ref{fig:max_cliques_decomp}. 

It can be inspected that the number $k$ of intervals in $I$ is equal to the number of peaks of the corresponding Dyck paths. The number of Dyck paths of length $2n$ with $k$ peaks is given by the Narayana number $N(n,k) = \frac{1}{n} {n \choose k} {n \choose k-1}$~\cite{Deutsch1999}. Note that this is what we obtained in Lemma~\ref{lemma:Cnk} in Appendix~\ref{app:nog}. Thus, the number of ordered graphs (and consequently of structures) with exactly $k$ maximal cliques is upper bounded by the Narayana number $N(n,k)$. 

Furthermore, it can also be inspected that the number of connected components of an ordered graph/structure is given by the number of returns to the horizontal axis of the corresponding Dyck path (including the last step). The number of Dyck paths of length $2n$ that touch the horizontal axis exactly $k+1$ times (including the endpoints) is equal to $\frac{k}{n} {2n-k-1 \choose n-1}$~\cite{Deutsch1999}, which is thus an upper limit to the number of ordered graphs and structures with exactly $k$ connected components.

Considering now connected ordered graphs, their maximal-clique representation must satisfy condition~\eqref{eq:cond_max_cliques} with the strict inequalities $a_{i+1} < b_i+1$, $1 \leq i \leq k-1$, such that no gaps exist between pairs of consecutive maximal cliques. The strict inequalities imply that the corresponding Dyck path never returns to zero (i.e., does not touch the horizontal axis) before the last step. Thus, the number of objects satisfying~\eqref{eq:cond_max_cliques} with the aforementioned strict inequalities is given by the number of Dyck paths that do not go below the horizontal line of height one. Since the first and last steps of a Dyck path of length $2n$ are always $u$ and $d$, respectively, the sought number of objects is therefore equal to the number of Dyck paths of length $2(n-1)$ (i.e., consisting of $2n-2$ steps), which is given by $c_{n-1}$. Alternatively, the same result can be obtained by requiring that the Dyck paths touch the horizontal axis just two times (at the start and end of the paths); setting $k=1$ in the expression from the previous paragraph, we get the number of admissible Dyck paths is equal to $\frac{1}{n} {2n-2 \choose n-1}\equiv c_{n-1}$.

\section{Proof of Proposition~\ref{prop:nog_exact}}\label{app:nog_exact}
To find the exact size of $\hat{\mathcal{G}}_{n,r}$, we use an alternative representation of an ordered graph based on partitioning the set of nodes into disjoint, nonempty blocks of nodes, each block being a clique in the ordered graph.

Specifically, for every $G \in \hat{\mathcal{G}}_{n,r}$ we define the blocks $\mathcal{B}_i = \mathcal{B}_i(G) = \{s_i,s_i+1,\ldots,t_i\}$, $i \geq 1$, where $s_1=1$,  $s_i = t_{i-1}+1$ for $i \geq 1$, and $t_i$ is the rightmost node connected by an edge to $s_i$ in $G$ (such that $s_i$ and $s_{i+1}$ are not connected). The blocks are disjoint and cover the set $\{1,\ldots,n\}$. Let $k \geq 1$ be the number of blocks given by $G$. We get $k=1$ when $G$ is the complete graph on $n$ nodes. For $k \geq 2$, the first nodes in consecutive blocks must be separated by at least $r$, because they are not connected. Therefore, the total range of the $n$ points is $X_{(n)}-X_{(1)} > (k-1)r$. At the same time, $X_{(n)}-X_{(1)} < 1$, since the points are distributed in $[0,1]$. This implies that the number of blocks given by any graph in $\hat{\mathcal{G}}_{n,r}$ must be $k \leq  \lceil 1/r \rceil$. Furthermore, denoting by $p_i$ be the number of nodes in the $i$th block ($p_i = |\mathcal{B}_i| = t_i-s_i+1$), we have $p_1+p_2+\ldots+p_k = n$ with $p_i \geq 1$, $i = 1,\ldots,k$. Thus, $k \leq \min\{ \lceil 1/r \rceil, n \}$.

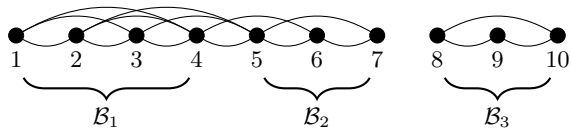
\begin{figure}
    \centering
    \def\d{0.8}
        \begin{tikzpicture}
            \footnotesize
            \node[fill,circle,inner sep=2pt,outer sep=0pt,draw,label=below:$1$] (N1)   at (0*\d,0) {};
            \node[fill,circle,inner sep=2pt,outer sep=0pt,draw,label=below:$2$] (N2)   at (1*\d,0) {};
            \node[fill,circle,inner sep=2pt,outer sep=0pt,draw,label=below:$3$] (N3)   at (2*\d,0) {};
            \node[fill,circle,inner sep=2pt,outer sep=0pt,draw,label=below:$4$] (N4)   at (3*\d,0) {};
            \node[fill,circle,inner sep=2pt,outer sep=0pt,draw,label=below:$5$] (N5)   at (4*\d,0) {};
            \node[fill,circle,inner sep=2pt,outer sep=0pt,draw,label=below:$6$] (N6)   at (5*\d,0) {};
            \node[fill,circle,inner sep=2pt,outer sep=0pt,draw,label=below:$7$] (N7)   at (6*\d,0) {};
            \node[fill,circle,inner sep=2pt,outer sep=0pt,draw,label=below:$8$] (N8)   at (7*\d,0) {};
            \node[fill,circle,inner sep=2pt,outer sep=0pt,draw,label=below:$9$] (N9)   at (8*\d,0) {};
            \node[fill,circle,inner sep=2pt,outer sep=0pt,draw,label=below:$10$](N10)  at (9*\d,0) {};
            \draw[bend right](N1) to node[]{} (N2);
            \draw[bend left] (N1) to node[]{} (N3);
            \draw[bend left] (N1) to node[]{} (N4);
            \draw[bend right](N2) to node[]{} (N3);
            \draw[bend left] (N2) to node[]{} (N4);
            \draw[bend left] (N2) to node[]{} (N5);
            \draw[bend right](N3) to node[]{} (N4);
            \draw[bend left] (N3) to node[]{} (N5);
            \draw[bend right](N4) to node[]{} (N5);
            \draw[bend left] (N4) to node[]{} (N6);
            \draw[bend right](N5) to node[]{} (N6);
            \draw[bend left] (N5) to node[]{} (N7);
            \draw[bend right](N6) to node[]{} (N7);
            \draw[bend right](N8) to node[]{} (N9);
            \draw[bend left](N8) to node[]{} (N10);
            \draw[bend right](N9) to node[]{} (N10);
            \draw [thick, decoration={brace,mirror,raise=16pt,amplitude=8pt},decorate] (N1) -- (N4) node [pos=0.5,anchor=north,yshift=-24pt] {$\mathcal{B}_1$};
            \draw [thick, decoration={brace,mirror,raise=16pt,amplitude=8pt},decorate] (N5) -- (N7) node [pos=0.5,anchor=north,yshift=-24pt] {$\mathcal{B}_2$};
            \draw [thick, decoration={brace,mirror,raise=16pt,amplitude=8pt},decorate] (N8) -- (N10) node [pos=0.5,anchor=north,yshift=-24pt] {$\mathcal{B}_3$};
    \end{tikzpicture}
    \caption{Partitioning the ordered graph of Fig.~\ref{fig:max_cliques_decomp} into disjoint blocks. Nodes in every block form a clique and can be connected by ``bridging'' edges only to nodes in adjacent blocks. There are no edges between blocks $\mathcal{B}_2$ and $\mathcal{B}_3$ because of the gap between nodes $7$ and $8$.}
    \label{fig:blocks}
\end{figure}

Each block constitutes a clique in the ordered graph. Furthermore, nodes in $\mathcal{B}_i$ may connect outside the block only to nodes in the adjacent blocks ($\mathcal{B}_{i-1}$ and $\mathcal{B}_{i+1}$). Let $B_{i-1,i}$ be the subgraph of $G$ that corresponds to the nodes in $\mathcal{B}_{i-1} \cup \mathcal{B}_i$. The graph $B_{i-1,i}$ inherits the property described in Lemma~\ref{lemma:OG_clique}. With this construction, every ordered graph in $\hat{\mathcal{G}}_{n,r}$ can be represented by the tuple $(B_{1,2},\ldots,B_{k-1,k})$. For an illustrative example, see Fig.~\ref{fig:blocks}. 

Conversely, we next show that to every proper tuple $(B_{1,2},\ldots,B_{k-1,k})$, $2 \leq k \leq  \lceil 1/r \rceil$, there corresponds a unique ordered graph from $\hat{\mathcal{G}}_{n,r}$. By ``proper'' we mean that the graphs of the tuple have the property given in Lemma~\ref{lemma:OG_clique}, and the leftmost and rightmost maximal cliques of $B_{i-1,i}$ are of sizes $p_{i-1}$ and $p_i$, for every $i \in \{2,\ldots, k \}$, with $p_1+p_2+\ldots+p_k = n$ and $p_1, \ldots, p_k \geq 1$. In this way, the blocks $\mathcal{B}_1,\ldots,\mathcal{B}_k$ can be identified and a graph with $n$ nodes can be constructed by piecing together $B_{1,2},\ldots,B_{k,k-1}$. Still, we need to show that for every proper tuple a suitable arrangement of the points in $[0,1]$ that gives a unique ordered graph exists. The case when $k=1$ (i.e., one single block) is straightforward, because this gives the complete graph on $n$ nodes, which is achieved by placing all $n$ points in an interval of length $<r$. For $k \geq 2$, it is sufficient to show that for any set of positions of the $p_{i-1}$ nodes of $\mathcal{B}_{i-1}$, we can place the $p_i$ nodes of $\mathcal{B}_i$ so that to achieve any desired proper graph $B_{i-1,i}$. This is because we can start with arbitrary locations for the nodes of the first block, $\mathcal{B}_1$, then place the nodes of $\mathcal{B}_2$ to obtain the graph $B_{1,2}$, and then successively place the nodes of subsequent block as specified by the tuple $(B_{1,2},\ldots,B_{k-1,k})$.

In more detail, let $x_{s_{i-1}},x_{s_{i-1}+1},\ldots,x_{t_{i-1}}$ be arbitrary locations of the $p_{i-1}$ nodes of $\mathcal{B}_{i-1}$, with $x_{t_{i-1}}-x_{s_{i-1}} < r$. We can view the given graph $B_{i-1,i}$ as basically encoding which of the nodes in $\mathcal{B}_i$ (if any) are connected to node $l \in \mathcal{B}_{i-1}$ but not to node $l-1$, for every $l \in \{s_{i-1}+1,\ldots,t_{i-1}\}$; those nodes are placed in the interval $(x_{l-1}+r, x_l+r)$. Note that by construction of the blocks, an edge cannot exist between any of the nodes in $\mathcal{B}_i$ and node $s_{i-1}$. Furthermore, in each step of this successive procedure, we can place the nodes arbitrarily close to each other and to the left end of their respective interval, such that the resulting length of the graph, which satisfies $x_n - x_1 > (k-1)r$, is arbitrarily close to $(k-1)r$.

Thus, we can count the number of ordered graphs in $\hat{\mathcal{G}}_{n,r}$ by counting the number of proper tuples $(B_{1,2},\ldots,B_{k-1,k})$. Given $p_{i-1}$ and $p_{i}$, the number of possible graphs $B_{i-1,i}$ is ${p_{i-1}+p_i-1 \choose p_i}$, as this is equivalent to counting the number of ways one can place $p_{i}$ balls in $p_{i-1}$ boxes (each box corresponding to an interval in which points may be placed, as described above). Since a graph $B_{i-1,i}$ can be achieved for any set of locations of the $p_{i-1}$ points of block $\mathcal{B}_{i-1}$, the number of tuples $(B_{1,2},\ldots,B_{k-1,k})$ for given $p_1,\ldots,p_k$ is $\prod_{i=2}^{k} {p_{i-1}+p_i-1 \choose p_i}$, $k \geq 2$. Thus, the number of ordered graphs with $k \geq 2$ blocks is 
\begin{equation}\label{eq:Tnk}
    T_{n,k} = \sum_{\substack{p_1,\ldots,p_k \geq 1\\ p_1+\ldots+p_k=n}}  \prod_{i=2}^{k} {p_{i-1}+p_i-1 \choose p_i}. 
\end{equation}
When $k=1$, $T_{n,1}=1$, since the one single block gives the complete graph in this case. 

Finally, the number of ordered graphs for a given connection range $r$ is
\begin{equation}\label{eq:Tn}
    |\hat{\mathcal{G}}_{n,r}| =  \sum_{k=1}^{\lceil 1/r \rceil} T_{n,k}
\end{equation}

In the following, we obtain an expression for $|\hat{\mathcal{G}}_{n,r}|$ in~\eqref{eq:Tn} by studying the generating function of the sequence $\{|\hat{\mathcal{G}}_{n,r}|\}_{n \geq 0}$,
\begin{align}
    g(x) &= \sum_{n=0}^\infty |\hat{\mathcal{G}}_{n,r}| x^n \\
    &=  \sum_{k=1}^{\lceil 1/r \rceil} g_k(x), \label{eq:g_gk}
\end{align}
where we used~\eqref{eq:Tn} and, for each $k\geq 1$, defined $g_k$ to be the generating function of $\{T_{n,k}\}_{n\geq 0}$, $g_k(x) = \sum_{n=0}^\infty T_{n,k} x^n$. For $k=1$, $T_{n,1}=1$, such that we have
\begin{equation}\label{eq:g1}
    g_1(x) = \sum_{n=0}^\infty x^n = \frac{1}{1-x}.
\end{equation}
For $k\geq 2$, we express $g_k$ using~\eqref{eq:Tnk} as
\begin{align}
    g_k(x) 
    &= \sum_{n=0}^\infty \sum_{\substack{p_1,\ldots,p_k \geq 1\\ p_1+\ldots+p_k=n}} \prod_{i=2}^{k} {p_{i-1}+p_i-1 \choose p_i} x^n \\
    &= \sum_{p_1,\ldots,p_k \geq 1}  \prod_{i=2}^{k} {p_{i-1}+p_i-1 \choose p_i} x^{p_1 + \ldots + p_k} \label{eq:g_sum_p}
\end{align}
The summation in~\eqref{eq:g_sum_p} can be extended such that its lower limit is zero by observing that whenever some $p_i = 0$, the binomial coefficient ${p_{i}+p_{i+1}-1 \choose p_{i+1}} = 0$, unless $p_{i+1}=0$ as well, meaning that all of the $p_{i+1},\ldots,p_k$ must be zero for the corresponding term in~\eqref{eq:g_sum_p} to be nonzero. Therefore, in~\eqref{eq:g_sum_p} we can sum over $p_1,\ldots,p_k \geq 0$ and then subtract the terms corresponding to $p_k=0$, which gives
\begin{equation}\label{eq:gk_dif}
    g_k(x) = f_k(x) - f_{k-1}(x), \quad k \geq 2,
\end{equation}
where we have defined
\begin{equation}
    f_k(x) = \sum_{p_1,\ldots,p_k \geq 0}  \prod_{i=2}^{k} {p_{i-1}+p_i-1 \choose p_i} x^{p_1 + \ldots + p_k}  
\end{equation}
We proceed to obtain an expression for $f_k$. Summing successively over the $p$'s,
\begin{align}
    f_k(x) 
    &= \sum_{p_1,\ldots,p_{k-1} \geq 0} \prod_{i=2}^{k-1} {p_{i-1}+p_i-1 \choose p_i} x^{p_1 + \ldots + p_{k-1}} \underbrace{\sum_{p_k \geq 0} {p_{k-1}+p_k-1 \choose p_k} x^{p_k}}_{=\frac{1}{\left(1-x\right)^{p_{k-1}}}} \\
    &= \sum_{p_1,\ldots,p_{k-2} \geq 0} \prod_{i=2}^{k-2} {p_{i-1}+p_i-1 \choose p_i} x^{p_1 + \ldots + p_{k-2}} \underbrace{\sum_{p_{k-1} \geq 0} {p_{k-2}+p_{k-1}-1 \choose p_{k-1}} \left(\frac{x}{1-x}\right)^{p_{k-1}}}_{=\frac{1}{\left(1-\frac{x}{1-x}\right)^{p_{k-2}}}}
\end{align}
Continuing in the same manner, we obtain an expression for $f_k(x)$ as a finite continued fraction with $k$ terms:
\begin{equation}
    f_k(x) = \cfrac{1}{ 1 - \cfrac{x}{ 1-\, _{{\ddots}_{\displaystyle -\cfrac{x}{1-x}}}}}.
\end{equation}
Thus, $f_k$ satisfies the recursion
\begin{equation}\label{eq:f_rec}
    f_0(x) = 1, \quad f_k(x) = \frac{1}{1-x f_{k-1}(x)}, \quad k\geq 1.
\end{equation}
The recursion~\eqref{eq:f_rec} corresponds to the generating function of the OEIS A080934 sequence~\cite{oeisA080934}; for example, this sequence gives the number of Dyck paths of length $2n$ and height less than or equal to $k$, and the number of rooted ordered trees on $n+1$ nodes of depth $\leq k+1$.

Finally, substituting~\eqref{eq:g1} and~\eqref{eq:gk_dif} into~\eqref{eq:g_gk}, we obtain that $g(x) = f_{\lceil 1/r \rceil}(x)$, i.e, the number of ordered graphs in $\hat{\mathcal{G}}_{n,r}$ is equal to the number of Dyck paths of length $2n$ and height at most $\lceil 1/r \rceil$, or, equivalently, the number of rooted ordered trees on $n+1$ nodes of depth $\leq \lceil 1/r \rceil+1$.

\section{Proof of Proposition~\ref{prop:ncs_LB}} \label{app:ncog_LB}
Using the block-based representation of an ordered graph introduced in Appendix~\ref{app:nog_exact}, the number of connected ordered graphs can be expressed as
\begin{equation}\label{eq:Tnc}
    |\hat{\mathcal{G}}_{n,r}^\text{c}| =  \sum_{k=1}^{\lceil 1/r \rceil} T_{n,k}^\text{c}
\end{equation}
where, similarly to~\eqref{eq:Tnk}, $T_{n,k}^\text{c}$ denotes the number of connected ordered graphs with $k\geq 2$ blocks, and is given by
\begin{equation}\label{eq:Tnkc}
    T_{n,k}^\text{c} = \sum_{\substack{p_1,\ldots,p_k \geq 1\\ p_1+\ldots+p_k=n}}  \prod_{i=2}^{k} \left[ {p_{i-1}+p_i-1 \choose p_i} - 1 \right]. 
\end{equation}
Compared to~\eqref{eq:Tnk}, the case where two consecutive blocks are disconnected is discounted in~\eqref{eq:Tnkc}. 

Using the approach of Appendix~\ref{app:nog_exact} to obtain an exact result leads to convoluted calculations, so we resort to establishing a lower bound on $|\hat{\mathcal{G}}_{n,r}^\text{c}|$. Specifically, for $T_{n,k}^\text{c}$ we only count the graphs where node $s_{i}$ (the first node of block $\mathcal{B}_{i}$) is connected to node $t_{i-1}$ (the last node of block $\mathcal{B}_{i-1}$) but not to $t_{i-1} - 1$. This count is essentially the number of ways one can place $p_i-1$ balls in $p_{i-1}-1$ boxes, such that we have 
\begin{equation}\label{eq:TnkcL}
    T_{n,k}^\text{c} > \sum_{\substack{p_1,\ldots,p_k \geq 1\\ p_1+\ldots+p_k=n}}  \prod_{i=2}^{k} {p_{i-1}+p_i-3 \choose p_i-1} = \tilde{T}_{n,k}^\text{c}. 
\end{equation}
We obtain the generating function of $\tilde{T}_{n,k}^\text{c}$ as follows:
\begin{align}
    \tilde{g}_k^\text{c}(x) 
     &= \sum_{n=0}^\infty \sum_{\substack{p_1,\ldots,p_k \geq 1\\ p_1+\ldots+p_k=n}} \prod_{i=2}^{k} {p_{i-1}+p_i-3 \choose p_i-1} x^n \\
    &= \sum_{p_1,\ldots,p_k \geq 1}  \prod_{i=2}^{k} {p_{i-1}+p_i-3 \choose p_i-1} x^{p_1 + \ldots + p_k}  \\
    &= x^k \sum_{p_1,\ldots,p_k \geq 0}  \prod_{i=2}^{k} {p_{i-1}+p_i-1 \choose p_i} x^{p_1 + \ldots + p_k}  \\
    &= x^k f_k(x),
\end{align}
where $f_k(x)$, given in Appendix~\ref{app:nog_exact}. Thus, $\tilde{T}_{n,k}^\text{c}$ is given by the coefficient of $x^{n-k}$ in the power series expansion of $f_k(x)$, i.e.,
\begin{equation}\label{eq:TnkcL_A}
    \tilde{T}_{n,k}^\text{c} = A_{n-k}(k),
\end{equation}
with $A$ given in~\eqref{eq:A}. Finally,~\eqref{eq:Tnc}, \eqref{eq:TnkcL} and~\eqref{eq:TnkcL_A} give the lower bound
\begin{equation}
    |\hat{\mathcal{G}}_{n,r}^\text{c}| >  \sum_{k=1}^{\lceil 1/r \rceil} A_{n-k}(k).
\end{equation}

\section{Proof of Proposition~\ref{prop:UBunif}} \label{app:proof_UBunif}

To evaluate the upper bound~\eqref{eq:H_UB_chain}, we need the joint distribution of the numbers of leftward neighbors nodes $i$ and $i+1$ have in the ordered graph. 

\subsection{Joint Distribution of $L_i$ and $L_{i+1}$}
For all $i\in\{2,\ldots,n-1\}$, we study $\Prob(L_i=a, L_{i+1}=b)$,  $0\leq a \leq i-1$ and $0 \leq b \leq a+1$. In the case where node $i$ is not connected by an edge to node $1$, i.e., $a<i-1$, we find that the joint probabilities do not depend on $i$. 

When $0 \leq a \leq i-2$ and $1 \leq b \leq a+1$, for nodes $i$ and $i+1$ to have $a$ and respectively $b$ leftward neighbours, they must be spatially located as follows:
$i-a-1$ points in $[0,X_{(i)}-r)$;
$a - b + 1$ points in $[X_{(i)}-r,X_{(i+1)}-r)$;
$b - 1$ points in $[X_{(i+1)}-r,X_{(i)})$;
$0$ points in $(X_{(i)},X_{(i+1)})$;
and, finally,  $n - i - 1$ points in $(X_{(i+1)},1]$. 
When the $n$ points are independent and uniformly distributed on $[0,1]$, we obtain the joint probability by integrating the multinomial distribution over the feasible values of the locations $x$ and $y$ of nodes $i$ and $i+1$, respectively. That is,  
\begin{equation}
    \Prob(L_i=a,L_{i+1}=b) = \iint\limits_{\mathcal{D}} \frac{ n! \, (x-r)^{i-a-1} (y-x)^{a-b+1} (r+x-y)^{b-1} (1-y)^{n-i-1} }{(i-a-1)! \, (a-b+1)! \, (b-1)! \, (n-i-1)! } \mathrm{d}x \, \mathrm{d}y,
\end{equation}
where the integration domain is $\mathcal{D} = \{(x,y)\in(0,1)^2 \mid x>r, y>x, y<x+r, y<1\}$. Assuming $r<1/2$, we make the change of variable $y-x=u$, such that $0<u<r$ and $r<x<1-u$, and obtain
\begin{align} \label{eq:Pab}
    \Prob(L_i=a,L_{i+1}=b) 
    &= \int_0^r \frac{ n! \, u^{a-b+1} (r-u)^{b-1}}{ (a-b+1)! \, (b-1)!} \nonumber \int_r^{1-u} \frac{  (x-r)^{i-a-1} (1-u-x)^{n-i-1} }{(i-a-1)!  \, (n-i-1)!} \mathrm{d}x \, \mathrm{d}u \nonumber\\
    &= \int_0^r \frac{ n! \, u^{a-b+1} (r-u)^{b-1} (1-r-u)^{n-a-1}}{ (a-b+1)! \, (b-1)! \, (n-a-1)!} \mathrm{d}u \nonumber \\
    &=  \int_0^1 \frac{ n! \, r^{a+1} \, t^{a-b+1} (1-t)^{b-1} (1-r-rt)^{n-a-1}}{ (a-b+1)! \, (b-1)! \, (n-a-1)!} \mathrm{d}t,
\end{align}
for all $0 \leq a \leq i-2$ and $1 \leq b \leq a+1$, and $r<1/2$. 

When $0 \leq a \leq i-2$ and $b =0$, the following must hold, assuming $r < 1/2$: $X_{(i+1)} - X_{(i)} > r$; $i-a-1$ points in $[0,X_{(i)}-r)$; $a$ points in $[X_{(i)}-r,X_{(i)})$; $0$ points in $(X_{(i)},X_{(i+1)})$; and, finally, $n - i - 1$ points in $(X_{(i+1)},1]$. Thus, we write
\begin{align} \label{eq:Pa0}
    \Prob(L_i=a,L_{i+1}=0) 
    &= \int_r^{1-r}  \int_{x+r}^1  \frac{ n! \, (x-r)^{i-a-1} r^a (1-y)^{n-i-1}}{(i-a-1)! \, a! \, (n-i-1)!} \, \mathrm{d}x \, \mathrm{d}y \nonumber \\
    &= {n \choose a} r^a (1-2r)^{n-a}.
\end{align}
Using~\eqref{eq:Pab} and~\eqref{eq:Pa0}, we find that
\begin{align}\label{eq:Pa}
    \Prob(L_i=a) &= \sum_{b=0}^{a+1} \Prob(L_i=a, L_{i+1}=b) \nonumber \\
    &= {n \choose a} r^a (1-r)^{n-a},
\end{align}
for all $a = 0,\ldots,i-2$. 

For $a=i-1$ (i.e., in the ordered graph, node $i$ is connected by an edge to node $1$), we therefore have 
\begin{align}\label{eq:Pi1}
    \Prob(L_i = i-1) &= 1 - \sum_{a=0}^{i-2} \Prob(L_i=a) \nonumber\\
    &= \sum_{a=i-1}^{n} {n \choose a} r^a (1-r)^{n-a}
\end{align}
Thus, $L_i$ has a truncated binomial distribution, as a consequence of the left margin. 

\subsection{The Normalized Entropy for Large $n$}
We consider the entropy per node, $\frac{1}{n}H(\hat{G}_{n,r})$. From~\eqref{eq:H_UB_chain}, we write
\begin{equation}\label{eq:H_UB_per_node}
    \frac{1}{n}H(\hat{G}_{n,r}) \leq \underbrace{\frac{1}{n} \sum_{i=2}^{n-1} \sum_{a=0}^{i-2} \Prob(L_i = a) H(L_{i+1} \mid L_i=a)}_{\equiv h_n} + \underbrace{\frac{1}{n} \sum_{i=2}^{n-1} \Prob(L_i = i-1) H(L_{i+1} \mid L_i = i-1)}_{\equiv c_n} + \frac{1}{n} H(L_2).
\end{equation}
The inner sum in the term in~\eqref{eq:H_UB_per_node} denoted by $h_n$ is taken over $a<i-1$. Consequently, the probabilities involved in $h_n$ (given by~\eqref{eq:Pab},~\eqref{eq:Pa0}, and~\eqref{eq:Pa}) do not depend on the node index $i$, such that we can express
\begin{equation}\label{eq:h_n}
    h_n = \frac{1}{n} \sum_{a=0}^{n-3} (n-a-2) \, \Prob(L_{n-1} = a) H(L_n \mid L_{n-1}=a).
\end{equation}

In the following we evaluate the normalized entropy in the large $n$ limit for different regimes of scaling of the connection range $r_n$.

We find it more convenient to work with $\Delta L_{n-1} = L_{n-1}+1 - L_n$ representing the difference between the indices of the leftmost nodes linked to nodes $n$ and $n-1$. Given $L_{n-1} = a$, $\Delta L_{n-1} \in \{0,\ldots,a+1 \}$. For example, $\Delta L_{n-1} = 0$ means nodes $n-1$ and $n$ have the same leftmost node, while $\Delta L_{n-1} = a+1$ indicates that node $n$ is isolated. Note that $H(L_n \mid L_{n-1}=a) =  H(\Delta L_{n-1} \mid L_{n-1}=a)$. Based on~\eqref{eq:Pab} and~\eqref{eq:Pa}, we find
\begin{equation}\label{eq:DL1}
    \Prob (\Delta L_{n-1} = k \mid L_{n-1} = a)  
    = \frac{(n-a)r_n}{(1-r_n)} {a \choose k} \int_0^1 t^k (1-t)^{a-k} \left(1 - \frac{r_n t}{1-r_n}\right)^{n-a-1} \mathrm{d}t
\end{equation}
for $k=0,\ldots,a$, whereas~\eqref{eq:Pa0} and~\eqref{eq:Pa} give
\begin{equation}\label{eq:DL2}
    \Prob (\Delta L_{n-1} = a+1 \mid L_{n-1} = a ) = \left(1- \frac{r_n}{1-r_n} \right)^{n-a}.
\end{equation}

\subsubsection{$r_n = O(1/n)$} 
We first consider $h_n$ given by~\eqref{eq:h_n}. $L_{n-1}$ has a binomial distribution\footnote{The truncation due to the left margin is negligible in this case, as $P(L_{n-1}=n-2)$ (i.e., the probability that the $(n-1)$th node is linked to node $1$, obtained from~\eqref{eq:Pi1}) is $O(n^{-n})$.} whose mean $n r_n$ is $O(1)$. Thus, the terms that have a significant contribution to the sum in~\eqref{eq:h_n}  correspond to small values of $a$. More specifically, for any $s>0$ and sequence $\epsilon_n > 0$, Markov's inequality gives
\begin{align}
    \Prob(L_{n-1} \geq n (r_n +\epsilon_n)) 
    &= \Prob(e^{s L_{n-1}} \geq e^{s n (r_n +\epsilon_n))} \nonumber\\
    &\leq e^{-s n (r_n +\epsilon_n)} \E\left[e^{s L_{n-1}}\right] \nonumber \\
    &= e^{-s n (r_n +\epsilon_n)} (1-r_n+r_n e^s)^n \nonumber \\
    &= e^{-s n \epsilon_n} e^{ -s n r_n + n \ln(1-r_n+r_n e^s) }
\end{align}
Choosing $\epsilon_n = \ln(n)/n$ and $s=2$, and given that $r_n = O(1/n)$, we obtain $\Prob(L_{n-1} \geq n r_n + \ln n)  = O(1/n^2)$. Furthermore, a rough bound on the entropy of $L_n$ based on the size of its support is $H(L_n \mid L_{n-1}=a) \leq \log_2(a+1) < \log_2 n$. Thus, the terms in~\eqref{eq:h_n} corresponding to $a \geq n r_n + \ln n = O(\ln n)$ have a contribution of $O(\log_2 n / n^2)$, which is a negligible quantity. 
We therefore estimate $H(L_n \mid L_{n-1}=a)$ for $a \ll n$. 

It can be shown that $\left(1 - \frac{r_n t}{1-r_n}\right)^{n-a-1} = e^{ -n r_n t} \left[ 1 + O(\ln n / n) \right]$, such that we obtain from~\eqref{eq:DL1} and~\eqref{eq:DL2} that, up to correction terms that vanish as $n$ grows large, $\Delta L_{n-1}$ has the conditional pmf
\begin{equation}\label{eq:DL_cpmf}
    \Prob (\Delta L_{n-1} = k \mid L_{n-1} = a) 
    \sim
    \begin{cases}
        \frac{n r_n}{a+1}  M(k+1,a+2,-n r_n), & k=0, \ldots, a, \\
        e^{-n r_n}, & k=a+1,
    \end{cases}
\end{equation}
where $M$ is Kummer's confluent hypergeometric function given by
\begin{equation*}
    M(\alpha,\beta,z) = \frac{\Gamma(\beta)}{\Gamma(\alpha) \, \Gamma(\beta-\alpha)} \int_0^1 t^{\alpha-1} (1-t)^{\beta-\alpha-1} e^{zt} \mathrm{d}t.
\end{equation*}
It can be verified that~\eqref{eq:DL_cpmf} is a proper pmf. Thus, we get
\begin{equation}
    H(L_n \mid L_{n-1}=a) =  - \sum_{k=0}^{a} \frac{n r_n}{a+1} M(k+1,a+2,-n r_n) \log_2\left(\frac{n r_n}{a+1} M(k+1,a+2,-n r_n)\right) + \frac{n r_n \, e^{-n r_n}}{\ln 2}.
\end{equation}
Since the binomial distribution of $L_{n-1}$ approaches the Poisson distribution with rate $n r_n$, the term $h_n$ given by~\eqref{eq:h_n} satisfies $h_n \sim h_\mathrm{U}(n r_n)$, where the function $h_\mathrm{U}$ is defined in~\eqref{eq:hU_func}.

We now consider the term $c_n$ in~\eqref{eq:H_UB_per_node}. The probability $\Prob(L_i = i-1)$ given by~\eqref{eq:Pi1} can be viewed as a tail of the binomial distribution. Similarly to the tail bound we obtained above for $L_{n-1}$, we can show that $\Prob(L_i = i-1) = O(1/n^2)$, for $i \geq n r_n + \ln n$. The entropy term satisfies $H(L_{i+1} \mid L_i = i-1) \leq \log_2 i<\log_2 n$. Therefore, 
\begin{align}
    c_n &= \frac{1}{n} \sum_{i=2}^{\lfloor n r_n + \ln n \rfloor} \Prob(L_i = i-1) H(L_{i+1} \mid L_i = i-1) \\
    &\phantom{=} + O \left( \frac{\log_2 n}{n^2} \right)\\
    &< \frac{1}{n} (n r_n + \ln n) \log_2(n r_n + \ln n) + O \left( \frac{\log_2 n}{n^2} \right) \\
    &= O \left( \frac{\ln n \ln\ln n}{n} \right).
\end{align}
Thus, $c_n$ is vanishing with $n$. 

The last term of~\eqref{eq:H_UB_per_node}, $H(L_2)/n$, is also vanishing, since $L_2$ is binary and therefore its entropy is no more than one bit. We can thus conclude that, when $r_n = O(1/n)$, the upper bound on the normalized entropy $\sim h_\mathrm{U}(n r_n)$.

\subsubsection{$r_n = \omega(1/n)$}
This implies $n r_n$ grows large with $n$. Examples include  a vanishing $r_n$ (e.g., $r_n \propto 1/\sqrt{n}$), or $r_n = r$, for some $r\in(0,1)$ bounded away from $0$ and $1$. 

Turning again to $h_n$ in~\eqref{eq:h_n}, in this case the sum is dominated by the terms corresponding to $a$ on the order of the mean $n r_n$ of $L_{n-1}$. For finite $k$ and $a$ large ($\sim n r_n$), we write~\eqref{eq:DL1} as
\begin{equation}
    \Prob (\Delta L_{n-1} = k \mid L_{n-1} = a) \sim \frac{(n-a)r_n}{(1-r_n)} \frac{a^k}{k!} \int_0^1 t^k \exp \left\{ a \ln(1-t) + (n-a) \ln \left(1 - \frac{r_n t}{1-r_n}\right) \right\} \mathrm{d}t.
\end{equation}
The argument of the exponential is strictly decreasing in $t$, and thus the integral is dominated by sufficiently small values of $t$. Linearizing the exponent,
\begin{align}
    \Prob (\Delta L_{n-1} = k \mid L_{n-1} = a)  
    & \sim \frac{(n-a)r_n}{(1-r_n)} \frac{a^k}{k!} \int_0^1 t^k e^{ - \left[ a + \frac{(n-a) r_n}{1-r_n} \right] t } \mathrm{d}t \\
    & \sim \frac{1}{1+\frac{a(1-r_n)}{(n-a)r_n}} \left( 1 - \frac{1}{1+\frac{a(1-r_n)}{(n-a)r_n}} \right)^k,
\end{align}
for $k=0,1,\ldots$. Thus, $\Delta L_{n-1} \mid L_{n-1} = a$ has a geometric distribution with success probability 
\begin{equation}
    p = \left( 1+\frac{a(1-r_n)}{(n-a)r_n} \right)^{-1},
\end{equation}
and its entropy is given by
\begin{equation}
    H(\Delta L_{n-1} \mid L_{n-1} = a) = -\log_2 p - \frac{1-p}{p} \log_2(1-p)
\end{equation}
To evaluate~\eqref{eq:h_n}, we express the entropy as a function of $x = \frac{a-nr_n}{(n-a)r_n}$, i.e., $H(\Delta L_{n-1} \mid L_{n-1} = a) = f(x)$ with $f:[-1,\infty)\to\mathbb{R}$,
\begin{equation}
    f(x) = (2+x)\log_2 (2+x) -(1+x)\log_2 (1+x).
\end{equation}
It can be verified that $f(0)=2$, $f'(0)=1$, and $f'(x)>0$ and $f''(x)<0$ over the domain of $f$ (i.e., $f$ is strictly increasing, strictly concave); therefore, $f(x)\leq 2+x$. Now, let $g(x) = f(x) - (2+x-x^2)$, which satisfies: $g(0)=0$, $g'(0)=0$ and $g''(0)>0$. Also, in addition to $x=0$, $g$ is also zero only at $x=-1$. Consequently, $g(x)\geq 0$, such that
\begin{equation}
    2+x(a)-x^2(a) \leq H(\Delta L_{n-1} \mid L_{n-1} = a) \leq 2+x(a).
\end{equation}
Using these inequalities in~\eqref{eq:h_n}, we obtain upper and lower limits on $h_n$ that differ by
\begin{align}
    \frac{1}{n} \sum_{a=0}^{n-3} \Prob(L_{n-1} = a) (n-a-2) \frac{(a-nr_n)^2}{(n-a)^2 r_n^2} 
    &= \frac{1}{n} \sum_{a=0}^{n-3} \frac{n! \, r_n^a (1-r_n)^{n-a}}{(n-a)! \, a!}  \frac{(a-nr_n)^2}{(n-a+1) r_n^2} \left[ 1 + O\left( \frac{1}{n} \right) \right] \\
    &= \frac{ \sum_{a=0}^{n-3} {n+1 \choose a} r_n^a (1-r_n)^{n-a+1} (a-nr_n)^2 }{n (n+1) r_n^2 (1-r_n)} \left[ 1 + O\left( \frac{1}{n} \right) \right] \\
    &= \left[ \frac{ (n+1) r_n (1-r_n) + r_n^2 }{n (n+1) r_n^2 (1-r_n)} + O\left(n^3 r_n^{n-4}\right) \right] \left[ 1 + O\left( \frac{1}{n} \right) \right] \\
    &=  \frac{1}{n r_n} + O\left( \frac{1}{n^2 r_n} \right) \xrightarrow{n\to\infty} 0
\end{align}
Thus, the upper and lower limits on $h_n$ are asymptotically equivalent to each other, and hence
\begin{align}
    h_n &\sim \frac{1}{n} \sum_{a=0}^{n-3} \Prob(L_{n-1} = a)  (n-a-2) \left[ 2 + \frac{(a-nr_n)}{(n-a) r_n} \right] \\
    &\sim 2(1-r_n),
\end{align}
where the result was obtained by taking similar steps to the above.

We now prove that $c_n$, defined in~\eqref{eq:H_UB_per_node} as
\begin{equation}\label{eq:cn2}
    c_n = \frac{1}{n} \sum_{i=2}^{n-1} \Prob(L_i = i-1) H(L_{i+1} \mid L_i = i-1)
\end{equation}
goes to zero as $n$ grows large. First, we show that the terms of the sum that defines $c_n$ are negligible for indices that satisfy $|i-n r_n-1|>\sqrt{n \ln n}$. Since $\Prob(L_i = i-1)$ given by~\eqref{eq:Pi1} is a tail probability, Hoeffding's inequality gives
\begin{equation}
    \Prob(L_i = i-1) \leq e^{ -2n \left( \frac{i-1}{n} - r_n \right)^2 }
\end{equation}
for all $i > n r_n + 1$. Therefore, 
\begin{equation}
    \Prob(L_i = i-1) \leq \frac{1}{n^2},
\end{equation}
for all $i > n r_n + 1 + \sqrt{n \ln n}$. Furthermore, the support of $L_{i+1} \mid L_i = i-1$ is $\{0,\ldots,i\}$, such that $H(L_{i+1} \mid L_i = i-1) \leq \log_2(i+1) \leq \log_2(n)$. Hence, the overall contribution of the terms in~\eqref{eq:cn2} corresponding to $i > n r_n + 1 + \sqrt{n \ln n}$ is $O(\log_2(n)/n^2)$, which is negligible for large $n$. Similarly, from~\eqref{eq:Pi1},
\begin{align}
    \Prob(L_{i+1} = i) 
    &= 1 - \sum_{a=0}^{i-1} {n \choose a} r^a (1-r)^{n-a} \\
    &\geq 1 - e^{ -2n \left( r_n - \frac{i-1}{n} \right)^2 } \\
    &\geq 1 - \frac{1}{n^2}
\end{align}
for all $i < n r_n + 1 - \sqrt{n \ln n}$. The probability that both the nodes $i$ and $i+1$ are connected by an edge to node 1 is the same as the probability that $i+1$ is linked to node 1, which translates to $\Prob(L_i=i-1,L_{i+1}=i) = \Prob(L_{i+1}=i)$, such that
\begin{equation}\label{eq:Plow}
    \Prob(L_{i+1}=i \mid L_i=i-1) = \frac{\Prob(L_{i+1}=i)}{\Prob(L_{i}=i-1)} \geq 1 - \frac{1}{n^2}.
\end{equation}
Therefore, using the log sum inequality and then~\eqref{eq:Plow},
\begin{align}
    H(L_{i+1} \mid L_i = i-1)  
    &= -\Prob(L_{i+1}=i \mid L_i=i-1) \log_2 \Prob(L_{i+1}=i \mid L_i=i-1)  \\
    &\phantom{=} -\sum_{k=0}^{i-1} \Prob(L_{i+1}=k \mid L_i=i-1) \log_2 \Prob(L_{i+1}=k \mid L_i=i-1) \\
    &\leq -\Prob(L_{i+1}=i \mid L_i=i-1) \log_2 \Prob(L_{i+1}=i \mid L_i=i-1) \\
    &\phantom{\leq} - \left[ 1 - \Prob(L_{i+1}=i \mid L_i=i-1) \right] \log_2 \left( \frac{ 1 - \Prob(L_{i+1}=i \mid L_i=i-1) }{i} \right)  \\
    &= O\left(\frac{\log_2 n}{n^2}\right),
\end{align}
for all $i < n r_n + 1 - \sqrt{n \ln n}$. 
Thus, the terms in~\eqref{eq:cn2} corresponding to $i < n r_n + 1 - \sqrt{n \ln n}$ have a negligible contribution. Finally, summing in~\eqref{eq:cn2} over the range $n r_n + 1 - \sqrt{n \ln n} < i < n r_n + 1 + \sqrt{n \ln n}$, and using $H(L_{i+1} \mid L_i = i-1) \leq \log_2(i+1) < \log_2(n r_n + 2 + \sqrt{n \ln n})$ and that the probabilities are less than one, we obtain
\begin{equation}
    c_n < \frac{1}{n} 2\lceil{\sqrt{n \ln n}}\rceil \log_2(n r_n + 2 + \sqrt{n \ln n}) \xrightarrow{n\to\infty} 0.
\end{equation}
Since $H(L_2)\leq 1$, we conclude that when $r_n = \omega(1/n)$ the upper bound on the entropy per node given by~\eqref{eq:H_UB_per_node} is asymptotically equivalent to $2(1-r_n)$.

\section{Proof of Proposition~\ref{prop:LBunif}} \label{app:proof_LBunif}

Based on~\eqref{eq:H_LB_chain}, we write
\begin{multline} \label{eq:H_LB_per_node}
    \frac{1}{n} H(\hat{G}_{n,r}) \geq \frac{1}{n} H(L_2) 
    + \underbrace{\frac{1}{n} \sum_{i=2}^{n-1} \sum_{a=0}^{i-2} \Prob(L_i=a) H\left(L_{i+1} \mid L_i=a, X_{(i-a)},\ldots,X_{(i)} \right)}_{h_n'} \\
    + \underbrace{\frac{1}{n} \sum_{i=2}^{n-1} \Prob(L_i=i-1) H\left(L_{i+1} \mid L_i=i-1, X_{(1)},\ldots,X_{(i)} \right)}_{c_n'},
\end{multline}
where we have isolated in the term denoted by $c_n'$ the case when $L_i=i-1$, i.e., the $i$th node is directly connected to the first node. 

In the following, we evaluate the term $h_n'$  in~\eqref{eq:H_LB_chain}. The variable $L_i$ has a truncated binomial distribution, see Appendix~\ref{app:proof_UBunif}. Focusing on the conditional distribution of $L_{i+1}$, let $\Delta_{(i+1)} = X_{(i+1)} - X_{(i)}$ be the gap between the $i$th and $(i+1)$th consecutive points. Given $L_i = a$, we define the distances $W_0,\ldots,W_{a-1}$ as $W_{a-b+1} = X_{(i-b+1)} - (X_{(i)}-r)$, $b = 2,\ldots,a+1$. The conditional distribution of $L_{i+1} \mid L_i = a$ is obtained by observing that:
\begin{itemize}
    \item $L_{i+1} = 0 \Leftrightarrow \Delta_{(i+1)} > r$; 
    \item $L_{i+1} = 1 \Leftrightarrow W_{a-1}<\Delta_{(i+1)} \leq r$; 
    \item $L_{i+1} = b \in \{2,\ldots,a\} \Leftrightarrow W_{a-b} < \Delta_{(i+1)} \leq W_{a-b+1}$;
    \item $L_{i+1} = a+1 \Leftrightarrow \Delta_{(i+1)} \leq W_0$.
\end{itemize}
When conditioning on $X_{(i)} = x > r$, the locations $X_{(i-a)}, \dots, X_{(i-1)}$ of the $a$ leftward neighbors of $i$ are independent of the $n-i$ locations $X_{(i+1)}, \ldots, X_{(n)}$; the former locations are i.u.d. on $[x-r,x]$, given $L_i$, whereas the latter are  i.u.d. over $[x,1]$. Therefore, the gap $\Delta_{(i+1)}$ is independent of $L_i$ given $X_{(i)}$ and its distribution is given by
\begin{equation}\label{eq:gap}
    \Prob\left(\Delta_{(i+1)} > u \mid X_{(i)}=x, L_i = a \right) = \left( 1 - \frac{u}{1-x} \right)^{n-i}, 
\end{equation}
where $0 \leq u \leq 1-x$. Furthermore,  given $L_i=a$, the distances $W_0,\ldots,W_{a-1}$ are independent of $X_{(i)}$, and their distribution is dictated by the fact that the $a$ points are i.u.d over an interval of length $r$. The following conditional pdfs are obtained: 
\begin{equation}\label{eq:pdfc_W0}
    f_{W_0 \mid L_i}(u \mid a) = \frac{a (r-u)^{a-1}}{r^a}, \quad 0<u<r,
\end{equation}
\begin{equation}
    f_{W_{a-1} \mid L_i}(u \mid a) = \frac{a u^{a-1}}{r^a}, \quad 0<u<r,
\end{equation}
and
\begin{equation}\label{eq:pdfc_WW}
    f_{W_{a-b},W_{a-b+1} \mid L_i}(u,v\mid a) = \frac{a! \, u^{a-b} (r-v)^{b-2}}{(a-b)! \, (b-2)! \, r^a},
\end{equation}
for $b=2,\ldots,a$ and $0<u<v<r$.

The joint distribution of $X_{(i)}$ and $L_i$ is characterized by
\begin{equation}\label{eq:PXL}
    \Prob\left( X_{(i)} \in [x,x+\mathrm{d}x], L_i = a \right) 
    = \frac{n! \, (x-r)^{i-a-1} \, r^a \, (1-x)^{n-i}}{(i-a-1)! \, a! \, (n-i)!} \mathrm{d}x,
\end{equation}
for $a \in \{0,\ldots,i-2\}$ and $x>r$.\footnote{Since $a \leq i-2$, such that the $i$th node is not connected by an edge to node $1$, it must hold that $x>r$, otherwise the two nodes are within the range $r$.} The probability~\eqref{eq:PXL} follows from the fact that the $n$ points are i.u.d. in $[0,1]$ and the desired event occurs when $i-a-1$ points fall in $[0,x-r]$, $a$ points in $(x-r,x)$, and $n-i$ points in $(x,1]$. Using~\eqref{eq:Pa}, we obtain the conditional density
\begin{equation}\label{eq:XcondL}
    f_{X_{(i)} \mid L_i=a}(x) = \frac{(n-a)! \, (x-r)^{i-a-1}  \, (1-x)^{n-i}}{ (1-r)^{n-a} (i-a-1)! \, (n-i)! },
\end{equation}
for $a \in \{0,\ldots,i-2\}$ and $x \in (r,1)$; the density is zero, otherwise. From~\eqref{eq:XcondL}, the conditional mean and variance of $X_{(i)}$ are $\mu = r + \frac{(1-r)(i-a)}{n-a+1}$ and $\sigma^2 = \frac{(1-r)^2(i-a)(n-i+1)}{(n-a+1)^2 (n-a+2)}$, respectively. For large $n$, the variance vanishes, such that, for any fixed $t$, we develop~\eqref{eq:gap} as
\begin{align}
    \Prob\left(\Delta_{(i+1)} > u \mid X_{(i)}=\mu+\sigma t, L_i = a \right) 
    &= \left( 1 - \frac{(n-a+1)u}{(1-r)(n-i+1)\left[1+O(1/\sqrt{n})\right]} \right)^{n-i} \nonumber\\
    &\sim e^{-\lambda_a u},
\end{align}
for all indices $i$ such that $n-i$ is large, and where $\lambda_a = \frac{n-a}{1-r}$. 

Based on the above, for all $i=2,\ldots, \lfloor n-\sqrt{n}\rfloor$ (such that $n-i$ is large), we express the conditional entropy as
\begin{equation}\label{eq:HcondX}
    H\left(L_{i+1} \mid L_i=a, X_{(i-a)},\ldots,X_{(i)} \right) \sim \sum_{b=0}^{a+1} g_{a,b}
\end{equation}
with $g_{a,0} =  -e^{-\lambda_a r} \log_2 e^{-\lambda_a r}$,
\begin{align*}
    g_{a,1} &=  \E \left[- \left( e^{-\lambda_a W_{a-1}} - e^{-\lambda_a r} \right) \log_2 \left( e^{-\lambda_a W_{a-1}} - e^{-\lambda_a r} \right)\right],\\
    g_{a,b} &=  \E \left[- \left( e^{-\lambda_a W_{a-b}} - e^{-\lambda_a W_{a-b+1}} \right) \log_2 \left( e^{-\lambda_a W_{a-b}} - e^{-\lambda_a W_{a-b+1}} \right)\right], \quad 2\leq b \leq a,\\
    g_{a,a+1} &=  \E\left[- \left( 1 - e^{-\lambda_a W_0} \right) \log_2 \left( 1 - e^{-\lambda_a W_0} \right)\right],
\end{align*}
where the expectations are with respect to the distances with conditional pdfs~\eqref{eq:pdfc_W0}--\eqref{eq:pdfc_WW}. By exchanging the expectation with the sum over $b$ and then evaluating the sum from $b=2$ to $a$, we obtain
\begin{align}
    \sum_{b=0}^{a+1} g_{a,b} &\sim \frac{\lambda_a r}{\ln 2} e^{-\lambda_a r} - \int_0^r \frac{a u^{a-1}}{r^a} \left( e^{-\lambda_a u} - e^{-\lambda_a r} \right) \log_2 \left( e^{-\lambda_a u} - e^{-\lambda_a r} \right) \mathrm{d}u \nonumber\\
    &\phantom{\sim} -\iint_{0<u<v<r} \frac{a(a-1)}{r^a} (r-v+u)^{a-2} \left(e^{-\lambda_a u} - e^{-\lambda_a v}\right) \log_2\left(e^{-\lambda_a u} - e^{-\lambda_a v}\right) \mathrm{d}u \, \mathrm{d}v \nonumber\\
    &\phantom{\sim} - \int_0^r \frac{a (r-v)^{a-1}}{r^a} \left( 1 - e^{-\lambda_a v} \right) \log_2 \left( 1 - e^{-\lambda_a v} \right) \mathrm{d}v
\end{align}
Given that for large $n$ \eqref{eq:HcondX} does not depend on $i$ in the range $2,\ldots,\lfloor n-\sqrt{n}\rfloor$, we truncate the sum that defines $h_n'$ in~\eqref{eq:H_LB_per_node} at $i=\lfloor n-\sqrt{n}\rfloor$. The error due to discarding the $\sqrt{n}$ terms vanishes with $n$, because the conditional entropy of $L_{i+1}$ is $O(\ln n)$ and the sum of probabilities is less than one, such that the truncation error is $O(\frac{\sqrt{n} \ln n}{n})$. Thus, we can express the double sum as a single sum over $a$, such that
\begin{equation}\label{eq:HL2}
    h_n' \sim \frac{1}{n} \sum_{a=0}^{\lfloor n-\sqrt{n}\rfloor - 2} (\lfloor n-\sqrt{n}\rfloor -a - 2) {n \choose a} r^a (1-r)^{n-a} \sum_{b=0}^{a+1} g_{a,b}.
\end{equation}

For large $n$, $\lambda_a \sim n$ and, considering $r=r_n$, \eqref{eq:HL2} becomes
\begin{align}\label{eq:HL3}
    h_n' &\sim (1-r_n)  \sum_{a=0}^n  {n \choose a} r_n^a (1-r_n)^{n-a} \sum_{b=0}^{a+1} g_{a,b} \nonumber \\
    &\sim \frac{1-r_n}{\ln 2} \bigg[ n r_n e^{-n r_n} - \int_0^{r_n} n (1-r_n+u)^{n-1} \left( e^{-n u} - e^{-n r_n} \right) \ln \left( e^{-n u} - e^{-n r_n} \right) \mathrm{d}u \nonumber\\
    &\phantom{\sim} -\iint_{0<u<v<r} n(n-1) (1-v+u)^{n-2} \left(e^{-n u} - e^{-n v}\right) \ln\left(e^{-n u} - e^{-n v}\right) \mathrm{d}u \, \mathrm{d}v \nonumber\\
    &\phantom{\sim} - \int_0^{r_n} n (1-v)^{n-1} \left( 1 - e^{-n v} \right) \ln \left( 1 - e^{-n v} \right) \mathrm{d}v \bigg]
\end{align}
Denote the three integral terms in~\eqref{eq:HL3} by $I_1$, $I_2$ and $I_3$ (in the order of their appearance). We obtain
\begin{align*}
    I_1 &\sim - \int_0^{r_n} n  e^{-n r_n} \left( 1 - e^{-n r_n} e^{n u} \right) \left[ -n u + \ln \left( 1 - e^{-n (r_n-u)} \right) \right] \mathrm{d}u \\
    &= n^2 r_n^2 e^{-n r_n} \int_0^1 \left( x - e^{-n r_n} x \, e^{n r_n x} \right) \, \mathrm{d}x - n r_n e^{-n r_n} \int_0^1 \left( 1 - e^{-n r_n x} \right) \ln \left( 1 - e^{-n r_n x} \right) \mathrm{d}x \\
    &= e^{-n r_n} \left( \frac{n^2 r_n^2}{2} - n r_n + 1 - e^{-n r_n}\right) - n r_n e^{-n r_n} \int_0^1 \left( 1 - e^{-n r_n x} \right) \ln \left( 1 - e^{-n r_n x} \right) \mathrm{d}x,
\end{align*}
where we substituted $x = (r_n-u)/r_n$. For $I_2$, we define $w = v-u$ and write the double integral as
\begin{equation*}
    I_2 \sim - \int_0^{r_n} \int_0^{r-w}  n^2 e^{-nw} e^{-nu} \left( 1 - e^{-n w} \right) \left[ -nu + \ln ( 1 - e^{-n w} \right] \mathrm{d}u \, \mathrm{d}w.
\end{equation*}
Integrating over $u$ and then over $w$ we finally obtain
\begin{equation*}
    I_2 \sim  \frac{1}{2}\left( 1-e^{-n r_n} \right)^2 - \frac{n^2 r_n^2}{2} e^{-n r_n} + n r_n e^{-n r_n} \int_0^1 \left( 1 - e^{-n r_n x} \right) \ln \left( 1 - e^{-n r_n x} \right) \mathrm{d}x + I_3,
\end{equation*}
where 
\begin{align*}
    I_3 &\sim - \int_0^{r_n} n  e^{-n v} \left( 1 - e^{-n v} \right) \ln \left( 1 - e^{-n v} \right) \mathrm{d}v \\
    &= - \int_0^{1-e^{-n r_n}} x \ln x \, \mathrm{d}x \\ 
    &= \frac{1}{4} \left( 1-e^{-n r_n} \right)^2 \left[ 1 - 2 \ln \left( 1-e^{-n r_n} \right) \right],
\end{align*}
where in the second line we made the substitution $x = 1-e^{-n v}$. Plugging the expressions for $I_1$, $I_2$, and $I_3$ back into~\eqref{eq:HL3}, we finally obtain that asymptotically in $n$,
\begin{equation}\label{eq:hnp}
    h_n' \sim (1-r_n) \left(1-e^{-n r_n}\right) \log_2 e   - (1-r_n) \left( 1-e^{-n r_n} \right)^2 \log_2 \left( 1-e^{-n r_n} \right)
\end{equation}
Given that conditioning reduces the entropy, the positive term denoted by $c_n'$ in~\eqref{eq:H_LB_per_node} is smaller than $c_n$ defined in~\eqref{eq:H_UB_per_node}. As shown in Appendix~\ref{app:proof_UBunif}, $c_n$ vanishes with $n$; therefore, $c_n'$ is also vanishing. Since $L_2$ is a binary variable, $H(L_2)/n$ also vanishes with $n$. Thus, the result~\eqref{eq:LBunif} is obtained by particularizing~\eqref{eq:hnp} for different scaling regimes of $r_n$.

\end{document}